\documentclass[prd,aps,nofootinbib,twocolumn,preprintnumbers]{revtex4}
\pdfoutput=1
\RequirePackage{snapshot} 
\usepackage{amsmath}
\usepackage{amssymb}
\usepackage{graphicx}
\usepackage{hyperref}
\usepackage{xspace}
\usepackage{wasysym} 
\usepackage{epstopdf}
\usepackage{booktabs}
\usepackage{hyperref}
\AtBeginDocument{
\heavyrulewidth=.08em
\lightrulewidth=.05em
\cmidrulewidth=.03em
\belowrulesep=.65ex
\belowbottomsep=0pt
\aboverulesep=.4ex
\abovetopsep=0pt
\cmidrulesep=\doublerulesep
\cmidrulekern=.5em
\defaultaddspace=.5em
}
\newcommand{\as}{\alpha_s}
\newcommand{\muR}{\mu_R}
\newcommand{\muF}{\mu_F}

\newcommand{\GeV}{\;\mathrm{GeV}}
\newcommand{\TeV}{\;\mathrm{TeV}}

\newcommand{\NNNLO}{\text{N$^3$LO}}

\usepackage{xcolor}
\definecolor{light-gray}{gray}{0.8}

\begin{document}

\title{Vector-Boson Fusion Higgs Pair Production at N$^3$LO}

\preprint{OUTP-18-09P, ZU-TH 38/18}

\newcommand{\OXaff}{Rudolf Peierls Centre for Theoretical Physics,
  University of Oxford,\\
  Clarendon Laboratory, Parks Road, Oxford OX1 3PU}

\newcommand{\UZHaff}{Department of Physics, University of Z{\"u}rich,
  CH-8057 Z{\"u}rich, Switzerland}

\author{Fr\'ed\'eric A. Dreyer}
\affiliation{\OXaff}
\author{Alexander Karlberg}
\affiliation{\UZHaff}

\begin{abstract}
  We calculate the next-to-next-to-next-to-leading order (\NNNLO) QCD
  corrections to vector-boson fusion (VBF) Higgs pair production in
  the limit in which there is no partonic exchange between the two
  protons.
  We show that the inclusive cross section receives negligible
  corrections at this order, while the scale variation uncertainties
  are reduced by a factor four.
  We present differential distributions for the transverse momentum
  and rapidity of the final state Higgs bosons, and show that there is
  almost no kinematic dependence to the third order corrections.
  %
  Finally we study the impact of deviations from the Standard Model in
  the trilinear Higgs coupling, and show that the structure of the
  higher order corrections does not depend on the self-coupling.
  These results are implemented in the latest release of the
  \texttt{proVBFH-incl} program.
\end{abstract}


\maketitle

\section{Introduction}
\label{sec:intro}
Following the discovery of the Higgs boson in
2012~\cite{Aad:2012tfa,Chatrchyan:2012xdj}, it has become a primary
focus of the experimental program of the Large Hadron Collider (LHC)
to measure its properties.
In particular, the measurement of the self-coupling of the Higgs boson
will be crucial both to further our understanding of the electroweak
symmetry breaking mechanism, and to constrain possible new physics
beyond the Standard Model (SM).

The simplest process with sensitivity to the trilinear Higgs coupling
at hadron colliders is the Higgs pair production process, which has
already been the focus of significant experimental
studies~\cite{Aaboud:2018sfw,Aaboud:2018ewm,Aaboud:2018ftw,Aaboud:2018knk,Aaboud:2016xco,Aad:2015xja,Aad:2015uka,Aad:2014yja,Sirunyan:2018tki,Sirunyan:2018iwt,Sirunyan:2017guj,Sirunyan:2017djm,Sirunyan:2017tqo}.
Due to the low cross sections, production rates at the LHC are very
small.  For this reason, processes with two final state Higgs bosons
are posed to play a key role at the high energy LHC (HE-LHC) and a
future 100 $\TeV$ circular collider (FCC) in probing the Higgs sector.
It is therefore important to have precise theoretical predictions for
the dominant channels.

As for single-Higgs production, the leading contribution at the
(HE-)LHC comes from gluon-gluon fusion~\cite{deFlorian:2016spz}.
This has been calculated up to next-to-next-to-leading order
(NNLO)~\cite{deFlorian:2013jea,deFlorian:2016uhr} matched to threshold
resummation at next-to-next-leading logarithmic (NNLL)
accuracy~\cite{deFlorian:2015moa}, and including finite top mass
effects~\cite{Grazzini:2018bsd}.

\begin{figure*}
  \centering
  \includegraphics[width=0.33\linewidth]{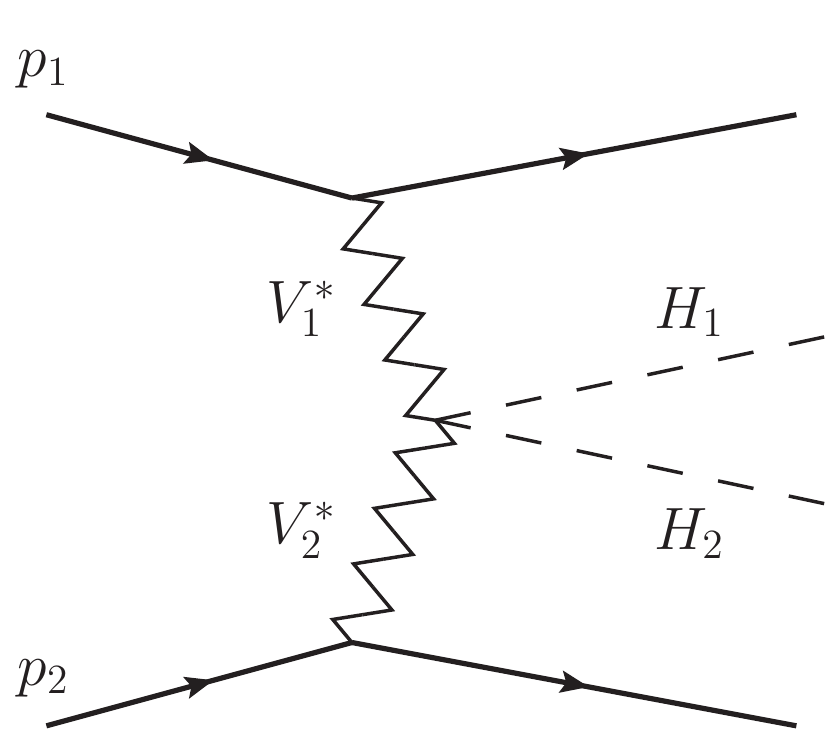}%
  \includegraphics[width=0.33\linewidth]{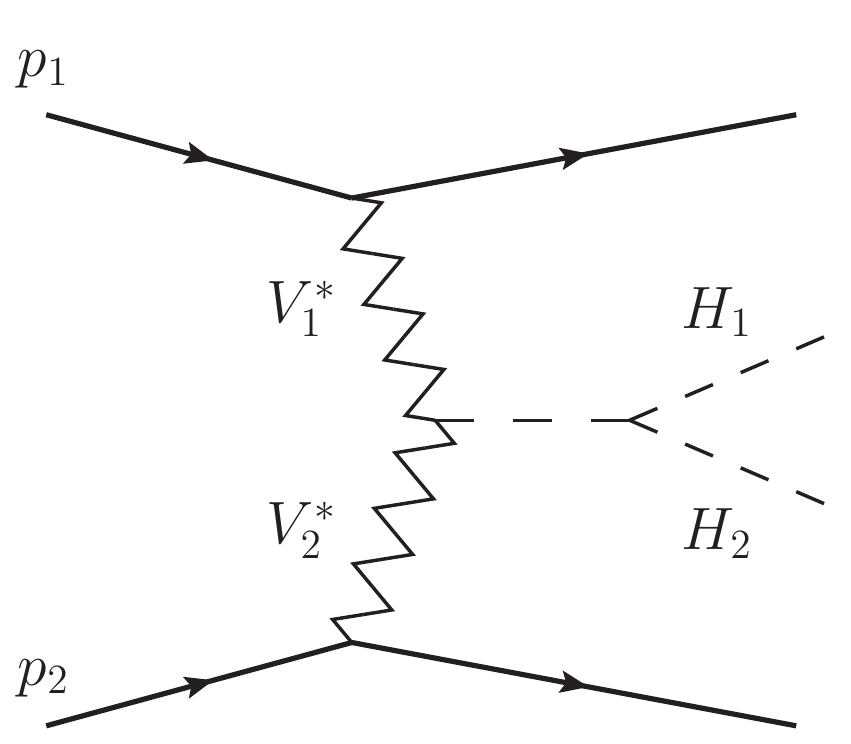}%
  \includegraphics[width=0.33\linewidth]{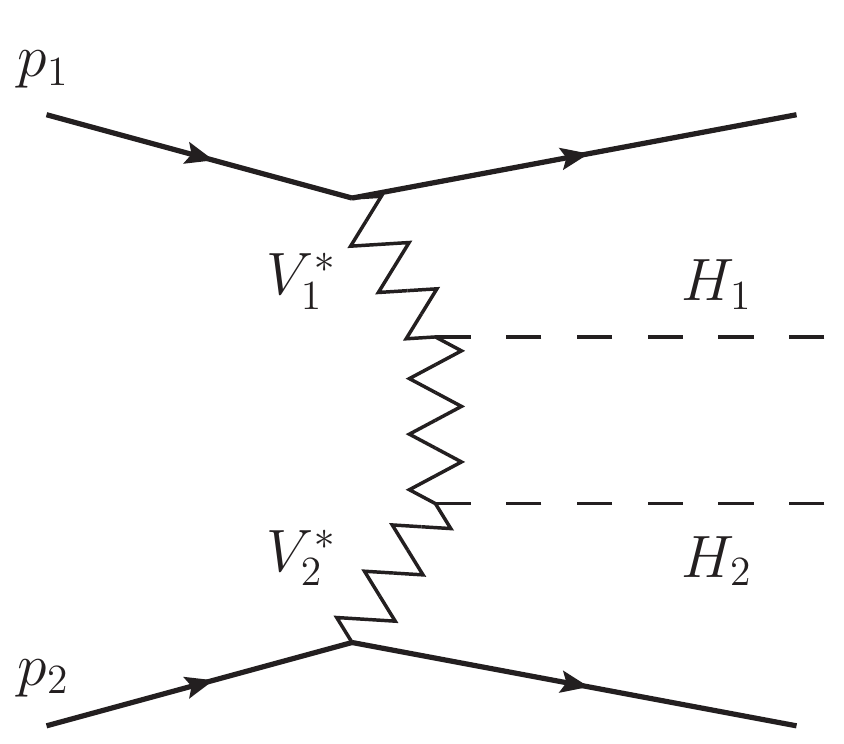}%
  \caption{Born-level diagrams contributing to VBF Higgs pair production.}
  \label{fig:vbfhh}
\end{figure*}

In this article, we focus on the vector-boson fusion (VBF) Higgs pair
production channel, shown at leading order in figure~\ref{fig:vbfhh}.
While it is only the second largest channel after gluon-gluon fusion,
the VBF production mode is one of particular interest for several
reasons: the presence of two tagging jets allows for a significant
reduction of the large backgrounds through an appropriate choice of
cuts; it is particularly sensitive to deviations from the SM in the
trilinear Higgs coupling~\cite{Baglio:2012np}; and is also a promising
channel for measurements of the $hhVV$ quartic coupling at the
LHC~\cite{Bishara:2016kjn}.

Because of the important role that double Higgs production via VBF
will play at the LHC and beyond, substantial efforts have been made to
calculate its cross section to high accuracy.
The differential cross section has been calculated up to
next-to-leading order (NLO)~\cite{Figy:2008zd,Baglio:2012np} with
matching to parton shower~\cite{Frederix:2014hta}, and up to
next-to-next-to-leading order when integrating out all hadronic final
states~\cite{Liu-Sheng:2014gxa}.

We present here the calculation of di-Higgs production up to
next-to-next-to-next-to-leading order (\NNNLO{}), which is also the
first calculation at this order of a $2\to 4$ process.
Together with a companion paper presenting the fully differential NNLO
calculation~\cite{Dreyer:2018rfu}, this brings the VBF double Higgs
channel to the same theoretical accuracy as single-Higgs VBF
production~\cite{Bolzoni:2010xr,Cacciari:2015jma,Dreyer:2016oyx,Cruz-Martinez:2018rod}.
These results are obtained using the structure function
approach~\cite{Han:1992hr}, which is the limit in which there is no
partonic exchange between the two protons, and in which all radiation
has been integrated over.
Since the single-gluon exchange is zero for color reasons, this
approximation is exact at NLO, while it has been shown to be accurate
to more than 1\% at NNLO for the single-Higgs
process~\cite{Ciccolini:2007ec,Harlander:2008xn,Bolzoni:2011cu}.
Because the presence of an additional Higgs boson does not impact the
color flow between the hadrons, this limit is expected to be just as
valid for Higgs pair production.

This paper is structured in the following way: In section~\ref{sec:hh}
we present the details of our calculation, in
section~\ref{sec:tot-xsc} we present results for the inclusive cross
section, while differential distributions are given in
section~\ref{sec:diff-dist}.
We give our conclusions in section~\ref{sec:conclusion}.
\section{Higgs pair production in VBF}
\label{sec:hh}

We start by setting up the formalism needed to calculate the inclusive
cross section up to third order in the expansion in the strong coupling
constant, which is analogous to the single-Higgs one.

The VBF Higgs pair production cross section is calculated as a double
deep inelastic scattering (DIS) process, and can be written
as~\cite{Han:1992hr}
\begin{align}
  \label{eq:vbfh-dsigma}
  d\sigma = &\,\sum_V\frac{G_F^2 m_V^4}{s}
  \Delta_V^2(Q_1^2)
  \Delta_V^2(Q_2^2) \,
  d\Omega_{\text{VBF}} \notag
  \\
  &\times
    \mathcal{W}^V_{\mu\nu}(x_1,Q_1^2)
    \mathcal{M}^{V,\mu\rho}
    \mathcal{M}^{V*,\nu\sigma}
  \mathcal{W}^{V}_{\rho\sigma}(x_2,Q_2^2)\,.
\end{align}
Here $G_F$ is Fermi's constant, $m_V$ and $\Delta_V^2$ are the mass
and squared propagators of the mediating $W$ or $Z$ bosons, and $\sqrt{s}$
is the collider center-of-mass energy.
We defined $Q_i^2=-q_i^2$ and $x_i = Q_i^2/(2 P_i\cdot q_i)$ as the
usual DIS variables, where $q_i$ is the four-momentum of the vector
boson $V_i$ and $P_i$ that of the initial proton.
Finally $\mathcal{W}^V_{\mu\nu}$ is the hadronic tensor and
$d\Omega_\text{VBF}$ is the four particle VBF phase space.
The matrix element of the $VV\to hh$ sub-process is expressed
as~\cite{Dobrovolskaya:1990kx} 
\begin{widetext}
\begin{multline}
  \label{eq:VV-subproc}
  \mathcal{M}^{V,\mu\nu} = 
  2 \sqrt{2} G_F g^{\mu\nu} \bigg(\frac{2m_V^4}{(q_1 + k_1)^2 - m_V^2 +i\Gamma_Vm_V}
  + \frac{2m_V^4 }{(q_1 + k_2)^2 - m_V^2+i\Gamma_Vm_V}
  + \frac{6 \nu \lambda m_V^2}{(k_1 + k_2)^2 - m_H^2+i\Gamma_Hm_H}
  + m_V^2 \bigg)
  \\
  + \frac{\sqrt{2} G_F m_V^4}{(q_1 + k_1)^2 - m_V^2}
  \frac{(2 k_1^\mu + q_1^\mu)(k_2^\nu - k_1^\nu - q_1^\nu)}{m_V^2-i\Gamma_Vm_V}
  + \frac{\sqrt{2} G_F m_V^4}{(q_1 + k_2)^2 - m_V^2}
  \frac{(2 k_2^\mu + q_1^\mu)(k_1^\nu - k_2^\nu - q_1^\nu)}{m_V^2-i\Gamma_Vm_V} \,,
\end{multline}
\end{widetext}
where $k_1, k_2$ are the final state Higgs momenta, which satisfy
$k_1+k_2 = q_1+q_2$, $\lambda$ is the trilinear Higgs self-coupling
and $\nu$ is the vacuum expectation value of the Higgs field.

Defining
$\hat{P}_{i,\mu} = P_{i,\mu} - \tfrac{P_i \cdot q_i}{q_i^2}
q_{i,\mu}$, the hadronic tensor $\mathcal{W}^V_{\mu\nu}$ in
equation~(\ref{eq:vbfh-dsigma}) is given by
\begin{multline}
  \label{eq:hadr-tensor}
  \mathcal{W}^V_{\mu\nu}(x_i,Q_i^2) = 
  \Big(-g_{\mu\nu}+\frac{q_{i,\mu}q_{i,\nu}}{q_i^2}\Big) F_1^V(x_i,Q_i^2)
  \\
  + \frac{\hat{P}_{i,\mu}\hat{P}_{i,\nu}}{P_i\cdot q_i} F_2^V(x_i,Q_i^2)
  + i\epsilon_{\mu\nu\rho\sigma}\frac{P_i^\rho q_i^\sigma}{2 P_i\cdot q_i} 
  F_3^V(x_i,Q_i^2)\,,
\end{multline}
where the $F^V_i(x,Q^2)$ functions are the standard DIS structure
functions with $i=1,2,3$, which can be expressed as a convolution of
the parton distribution functions (PDF) with the short distance
coefficient functions
\begin{equation}
  \label{eq:conv-structf}
  F_i^V = \sum_{a=q,g} C_i^{V,a} \otimes f_a \,,\quad i=1,2,3\,.
\end{equation}
To evaluate equation~(\ref{eq:conv-structf}), it is useful to define
the singlet and non-singlet distributions
$q_\text{S}\,, q_{\text{NS},i}$, as well as the non-singlet valence
distribution $q_\text{NS}^V$ and the asymmetry $\delta q_{\text{NS}}^{\pm}$
\begin{gather}
  q_\text{S}\hspace{-0.5mm} =\hspace{-0.5mm} \sum_{j=1}^{n_f}(q_j\hspace{-0.5mm} +\hspace{-0.5mm} \bar{q}_j),\;\;
  q_{\text{NS},j}^\pm \hspace{-0.5mm}=\hspace{-0.5mm} q_j \hspace{-0.5mm}\pm\hspace{-0.5mm} \bar{q}_j,\;\;
  q_\text{NS}^v \hspace{-0.5mm}=\hspace{-0.5mm} \sum_{j=1}^{n_f} (q_j\hspace{-0.5mm} -\hspace{-0.5mm} \bar{q}_j),\nonumber\\
  \delta q_{\text{NS}}^{\pm} =
  \sum_{u\text{-type}}(q_j \pm \bar{q}_j)
  - \sum_{d\text{-type}}(q_j \pm \bar{q}_j)\,.
\end{gather}
We can then decompose the quark coefficient functions into non-singlet
and pure-singlet parts, and define the valence coefficient function
\begin{gather}
  \label{eq:valence-coef}
  C_{L,q} = C^+_{L,\text{NS}} + C_{L,\text{PS}}\,,\quad
  C_{2,q} = C^+_{2,\text{NS}} + C_{2,\text{PS}}\,,\nonumber\\
  C^v_{3,\text{NS}} =   C^-_{3,\text{NS}} + C^s_{3,\text{NS}}\,,
\end{gather}
The neutral current structure functions can now be expressed as
\begin{multline}
  \label{eq:structf-Z-2L}
  F_i^Z(x) = 
  2x\hspace{-0.5mm}\int_0^1\hspace{-1.2mm} dz \int_0^1\hspace{-1.2mm}
  dy \delta(x \hspace{-0.3mm}-\hspace{-0.3mm} yz) 
  \sum_{j=1}^{n_f}\big[(v^{Z}_j)^2 + (a^Z_j)^2\big]
  \\
  \times\hspace{-1.4mm}\Big[
  q^+_{\text{NS},j}\hspace{-0.1mm}(\hspace{-0.1mm}y\hspace{-0.1mm})
  C_{i,\text{NS}}^+\hspace{-0.1mm}(\hspace{-0.1mm}z\hspace{-0.1mm})
  \hspace{-0.5mm}+\hspace{-0.5mm}
  q_\text{S}\hspace{-0.1mm}(\hspace{-0.1mm}y\hspace{-0.1mm})
  C_{i,\text{PS}}\hspace{-0.1mm}(\hspace{-0.1mm}z\hspace{-0.1mm})
  \hspace{-0.5mm}+\hspace{-0.5mm}
  g\hspace{-0.1mm}(\hspace{-0.1mm}y\hspace{-0.1mm})
  C_{i,g}\hspace{-0.1mm}(\hspace{-0.1mm}z\hspace{-0.1mm})
  \Big],\hspace{-2.5mm}
\end{multline}
\begin{multline}
  \label{eq:structf-Z-3}
  F_3^Z(x) = 4 \int_0^1 dz \int_0^1 dy \delta(x - yz)
  \sum_{j=1}^{n_f} v^Z_j a^Z_j
  \\
  \times
  \Big[
    q_{\text{NS},j}^-(y) C^-_{3,\text{NS}}(z)
    + q_{\text{NS}}^v(y) C_{3,\text{NS}}^v(z)
  \Big]\,,
\end{multline}
where $i=2,L$ and $F_1^Z = \tfrac1{2x}(F_2^Z - F_L^Z)$.
The vector and axial-vector coupling constants $v^Z_i$ and $a^Z_i$ are
given by 
\begin{align}
  \label{eq:av-couplings-Z}
  v^Z_j = 
  \pm\frac{1}{2}\,,\quad
  a^Z_j = 
       \left\{\begin{array}{cc}
           \frac{1}{2} - \frac{4}{3}\sin^2\theta_w, & u\text{-type}
           \\
           -\frac{1}{2} + \frac{2}{3}\sin^2\theta_w, & d\text{-type}
         \end{array}\right.
\end{align}
For the charged current case, the structure functions can be written as
\begin{multline}
  \label{eq:structf-W-2L}
  F_i^{W^\pm} \hspace{-0.4mm}(x) \hspace{-0.3mm} =\hspace{-0.3mm}
  \frac{x}{n_f}\hspace{-0.8mm}
  \int_0^1\hspace{-1.5mm} dz \hspace{-0.7mm}\int_0^1\hspace{-1.5mm}
  dy \delta(x\hspace{-0.3mm} - \hspace{-0.3mm}yz)
  \hspace{-0.7mm}
  \sum_{j=1}^{n_f}\hspace{-0.3mm}
  \big[(v^W_j)^2 \hspace{-0.7mm}+\hspace{-0.4mm} (a^W_j)^2\big]
  \\
  \times\hspace{-1.4mm}\Big[\hspace{-0.5mm}
  \mp\hspace{-0.5mm}
  \delta q_{\text{NS}}^{-}\hspace{-0.1mm}(\hspace{-0.1mm}y\hspace{-0.1mm})
  C_{i,\text{NS}}^{-}\hspace{-0.1mm}(\hspace{-0.1mm}z\hspace{-0.1mm})
  \hspace{-0.5mm}+ \hspace{-0.5mm}
  q_\text{S}\hspace{-0.1mm}(\hspace{-0.1mm}y\hspace{-0.1mm})
  C_{i,q}\hspace{-0.1mm}(\hspace{-0.1mm}z\hspace{-0.1mm})
  \hspace{-0.5mm}+\hspace{-0.5mm}
  g\hspace{-0.1mm}(\hspace{-0.1mm}y\hspace{-0.1mm})
  C_{i,g}\hspace{-0.1mm}(\hspace{-0.1mm}z\hspace{-0.1mm})
  \Big],\hspace{-2.5mm}
\end{multline}
\begin{multline}
  \label{eq:structf-W-3}
  F_3^{W^\pm}(x) = \frac{2}{n_f} \int_0^1 dz \int_0^1 dy \delta(x - yz)
  \sum_{j=1}^{n_f} v^W_j a^W_j
  \\
  \times
  \Big[
    \mp\delta q_{\text{NS}}^+(y) C^+_{3,\text{NS}}(z)
    + q_{\text{NS}}^v(y) C_{3,\text{NS}}^v(z)
  \Big]\,,
\end{multline}
where we have again
$F_1^{W^\pm} = \tfrac1{2x}(F_2^{W^\pm} - F_L^{W^\pm})$, and the
couplings are simply
$a^W_j = v^W_j = \tfrac{1}{\sqrt{2}}$.

\begin{figure}
  \centering
  \includegraphics[width=0.5\linewidth]{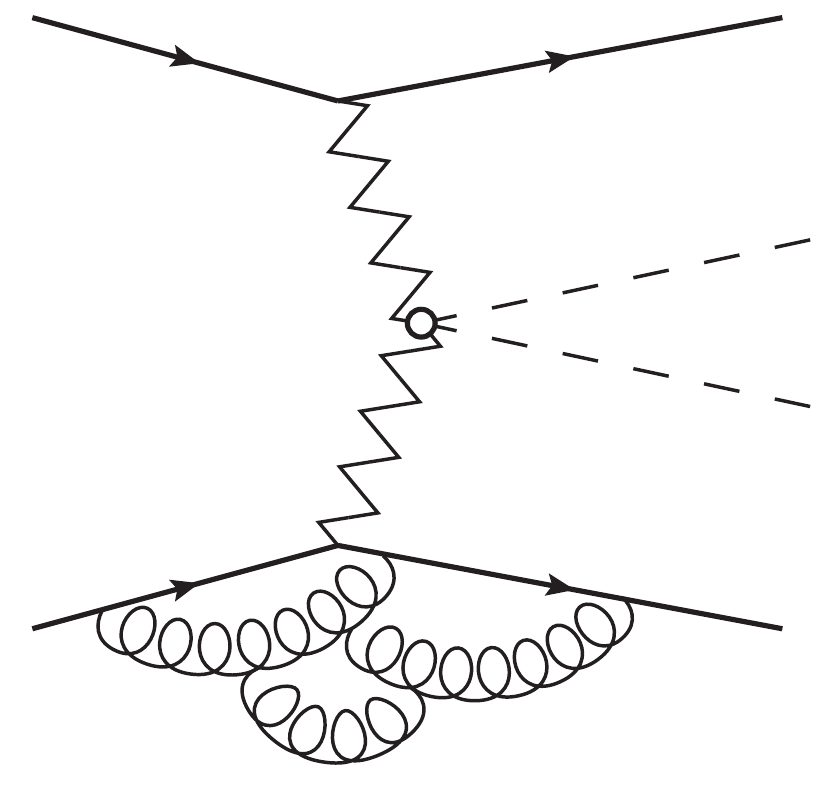}%
  \includegraphics[width=0.5\linewidth]{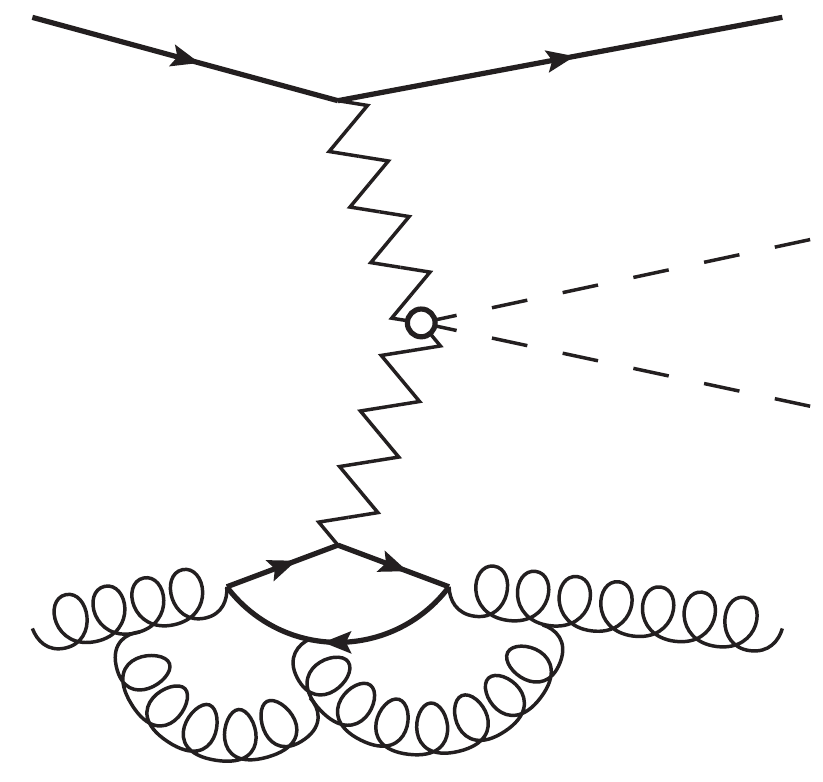}%
  \caption{Three-loop diagrams contributing at \NNNLO{} to VBF Higgs
    pair production.}
  \label{fig:vbfhh-as3}
\end{figure}

We can calculate corrections up to $\NNNLO{}$ by making use of the
known three-loop coefficient
functions~\cite{Moch:2004xu,Vermaseren:2005qc,Vogt:2006bt,Moch:2007rq,Davies:2016ruz},
whose parameterized expressions have been implemented
in \texttt{HOPPET} v1.2.0-struct-func-devel~\cite{Salam:2008qg}.
Examples of three-loop diagrams included in this calculation are shown
in figure~\ref{fig:vbfhh-as3}.

To calculate the variation of the cross section with different choices of
factorization and renormalization scales, we compute the scale dependence to
third order in the coefficient functions as well as in the PDFs.
\begin{figure}
  \centering
          \includegraphics[clip,width=0.45\textwidth,page=1,angle=0]{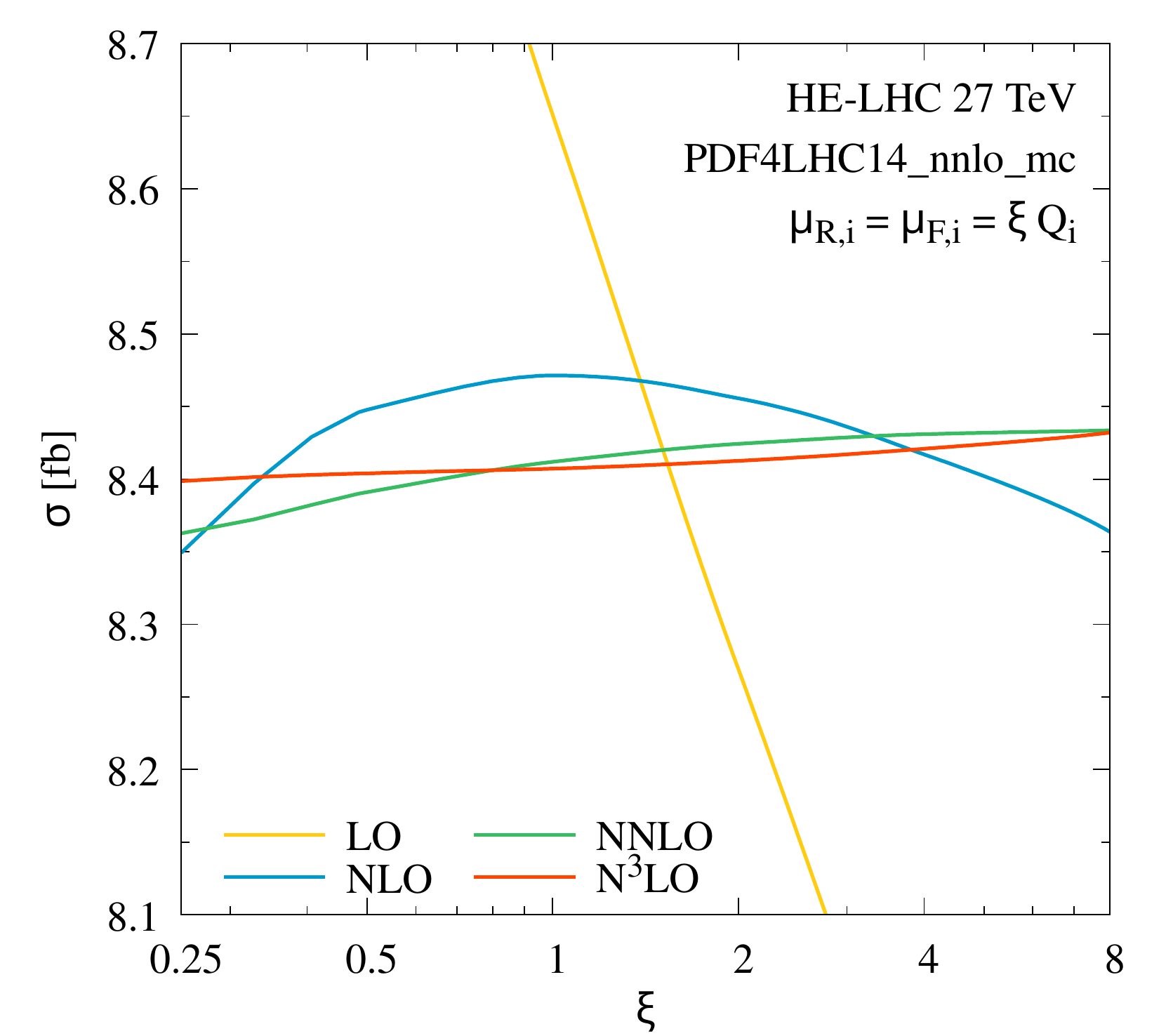}
          \caption{Total cross section as a function of the
            renormalization and factorization scales for each order in
            the perturbative expansion.}
  \label{fig:total-cross-sections-var}
\end{figure}

We start by evaluating the running coupling for $\as$
\begin{multline}
  \label{eq:as-running}
  \as(Q) \simeq \as(\muR) + \as^2(\muR) \beta_0 \ln\Big(\frac{\muR^2}{Q^2}\Big) \\
  + \as^3(\muR) \bigg[\beta_0^2 \ln^2\Big(\frac{\muR^2}{Q^2}\Big)
     + \beta_1 \ln\Big(\frac{\muR^2}{Q^2}\Big)\bigg]\,,
\end{multline}
where we defined $\beta_0 = (33 - 2 n_f)/12\pi$ and
$\beta_1 = (153 - 19 n_f)/24\pi^2$.
The coefficient functions can then easily be expressed as an expansion
in $\as(\mu_R)$.
To evaluate the dependence of the PDFs on the factorization scale
$\muF$, we can integrate the DGLAP equation, using
\begin{equation}
  \label{eq:pdf-integ}
  f(x,Q) = f(x,\muF) - \int_0^{L_{FQ}} dL \frac{d}{dL} f(x,\mu) \,.
\end{equation}
Expressing the PDF in terms of an expansion in $\as(\muR)$ evaluated
at $\muF$, it then straightforward to evaluate
equation~(\ref{eq:conv-structf}) for any choice of the renormalization
and factorization scales up to \NNNLO{}.

To estimate the theoretical uncertainty due to missing higher order
corrections, we calculate the envelope of seven different scale
choices, taking
\begin{equation}
  \label{eq:scale-var}
  \mu_R = \xi_R\, \mu_0\,,\quad \mu_F = \xi_F\, \mu_0\,,\quad
  \xi_{R,F} \in \{\tfrac12,1,2\}\,,
\end{equation}
where we keep $\tfrac12 \leq \mu_R/\mu_F \leq 2$ and $\mu_0$ is the
central scale choice.
We set the central renormalization and factorization scales to the
vector boson virtuality of the corresponding sector, $Q_1$ or $Q_2$.

For the numerical integration, we use the phase space parameterization
of \texttt{VBFNLO}~\cite{Arnold:2011wj}.
Unless otherwise specified the center-of-mass energy is set to the
expected energy of the HE-LHC, which is 27 TeV.
For all simulations, we use the \texttt{PDF4LHC15\_nnlo\_mc} 
set~\cite{Butterworth:2015oua} with a four-loop evolution of the
strong coupling, starting from an initial condition $\as(M_Z) = 0.118$.
We set the mass of the Higgs boson to $m_H = 125 \GeV$.
The electroweak parameters are set to the PDG
values~\cite{Tanabashi:2018oca}, with $m_W = 80.379 \GeV$,
$m_Z = 91.1876 \GeV$ and $G_F=1.16637 \cdot 10^{-5} \GeV^{-2}$.
The narrow-width approximation is used for the final state Higgs
bosons, while Breit-Wigner distributions are used for internal bosons,
taking $\Gamma_W = 2.085 \GeV$, $\Gamma_Z = 2.4952 \GeV$, and
$\Gamma_H=4.030\cdot 10^{-3} \GeV$.

\begin{figure}
  \centering
  \includegraphics[clip,width=0.45\textwidth,page=1,angle=0]{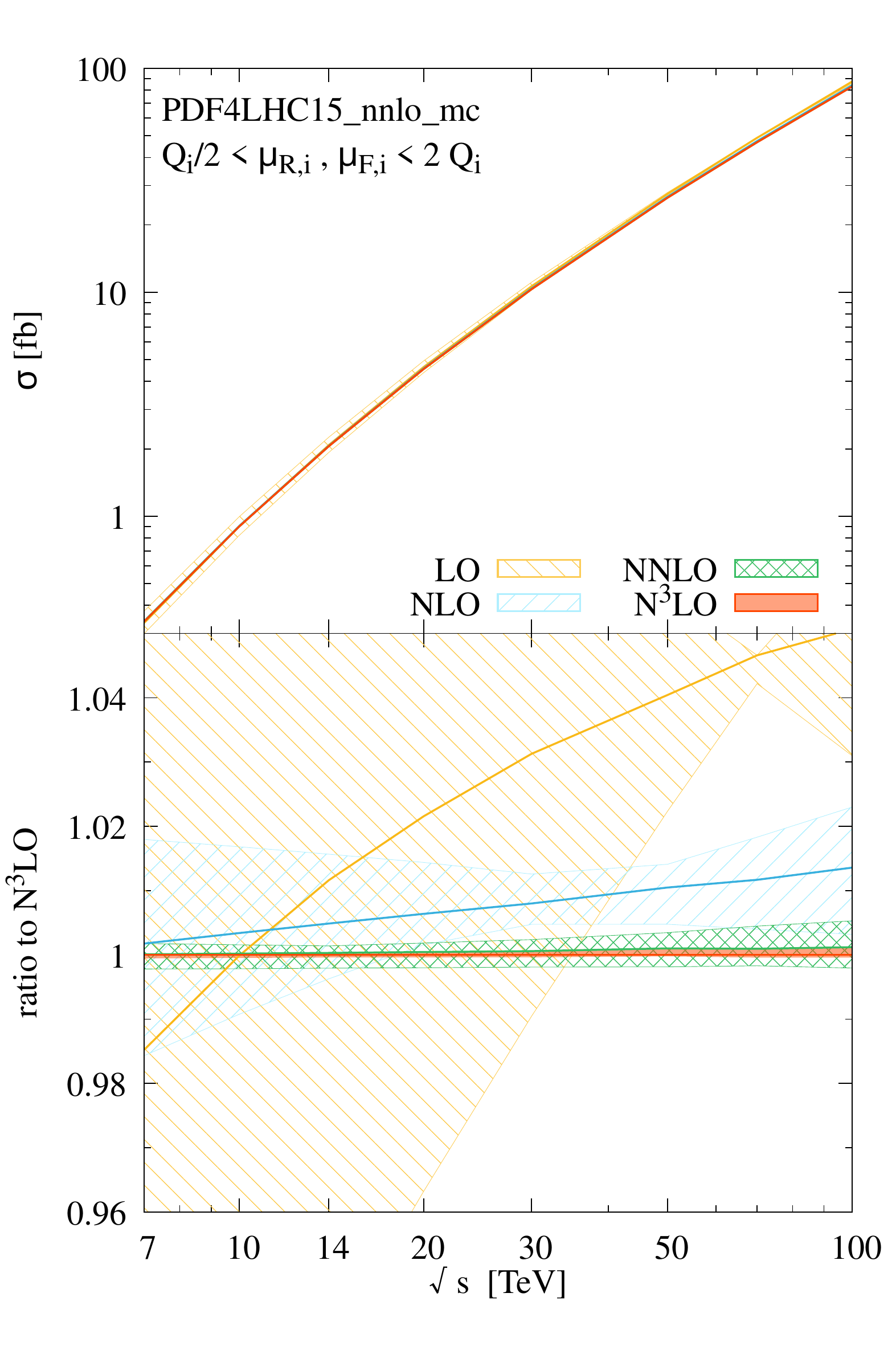}
  \caption{Total cross section as a function of energy for
    each order in perturbative QCD.}
  \label{fig:total-cross-sections-sqrts}
\end{figure}

\section{Total cross section}
\label{sec:tot-xsc}

We start by providing results for the inclusive cross section.

In figure~\ref{fig:total-cross-sections-var}, we show the dependence
of the total cross section on the renormalization and factorization
scales for each order in QCD.
One can clearly observe the convergence of the perturbative expansion,
with each order in $\as$ reducing the fluctuations due to changes in
the choice of scale.
We see that at \NNNLO{} there is almost no residual dependence on the
scale, with predictions having an almost constant cross section over a
broad range of scale values.

We show the dependence of the total cross section as a function of
center-of-mass energy in figure~\ref{fig:total-cross-sections-sqrts}.
Here we see that at even very high energies, the third order
corrections are fully contained within the NNLO scale variation bands,
with an almost constant $K$-factor.
One should note that this is somewhat dependent on the choice of
central scale, and less dynamical scales such as an $m_h$ or
$p_{t,HH}$ based prescription will lead to third order corrections
that can deviate from the NNLO uncertainty bands in certain kinematic
regions or at sufficiently high energies.

\begin{table}[t] 
  \centering
  \phantom{x}\medskip
  \begin{tabular}{lcccccc}
    \toprule
    && $\sigma^\text{(14 TeV)}$ [fb] && $\sigma^\text{(27 TeV)}$ [fb] && $\sigma^\text{(100 TeV)}$ [fb] \\
    \midrule
    LO       && $2.079\,^{+0.177}_{-0.152}$ && $8.651\,^{+0.411}_{-0.382}$ && $87.104\,^{+1.023}_{-1.633}$\\[4pt]
    NLO      && $2.065\,^{+0.022}_{-0.018}$ && $8.471\,^{+0.046}_{-0.024}$ && $84.026\,^{+0.781}_{-0.860}$\\[4pt]
    NNLO     && $2.056\,^{+0.003}_{-0.005}$ && $8.412\,^{+0.014}_{-0.021}$ && $83.000\,^{+0.340}_{-0.269}$\\[4pt]
    \NNNLO{} && $2.055\,^{+0.001}_{-0.001}$ && $8.407\,^{+0.005}_{-0.003}$ && $82.901\,^{+0.097}_{-0.035}$\\
    \bottomrule
  \end{tabular}
  \caption{Total cross sections at LO, NLO, NNLO and \NNNLO{} for VBF Higgs
    pair production for different center-of-mass energies.
    The uncertainties are obtained from a seven-point scale
    variation.}
\label{tab:cross-sections}
\end{table}

We detail the precise value of the cross section and its scale
variation uncertainties in table~\ref{tab:cross-sections}.
Values are given for three reference center-of-mass energies: the 14
TeV LHC, the 27 TeV HE-LHC and the 100 TeV FCC.
For each of these energies, we provide inclusive cross sections at
each order in perturbative QCD, along with the corresponding scale
variation envelope.
We can observe that while the corrections are at the level of a few
permille only, the scale uncertainty bands are reduced by more than a
factor four when going from NNLO to \NNNLO{}.

A comment is due on the impact of contributions beyond those included
in the DIS limit.
There are a number of corrections to the Born diagrams shown
in figure~\ref{fig:vbfhh} beyond those due to the radiation of
additional partons.
These should be included where possible for precise phenomenological
predictions.

\begin{figure*}
  \centering
  \includegraphics[clip,width=0.33\textwidth,page=2,angle=0]{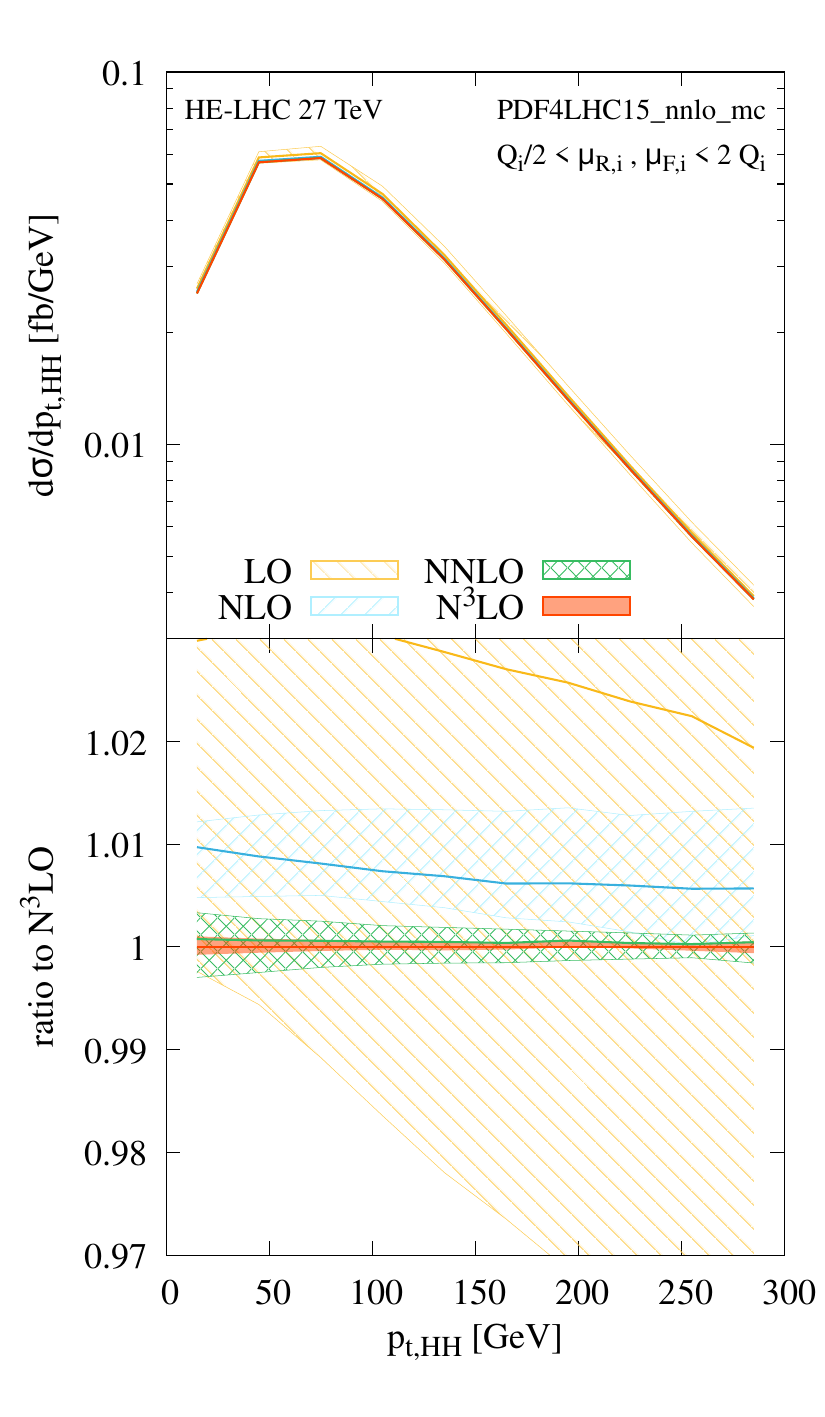}%
  \includegraphics[clip,width=0.33\textwidth,page=5,angle=0]{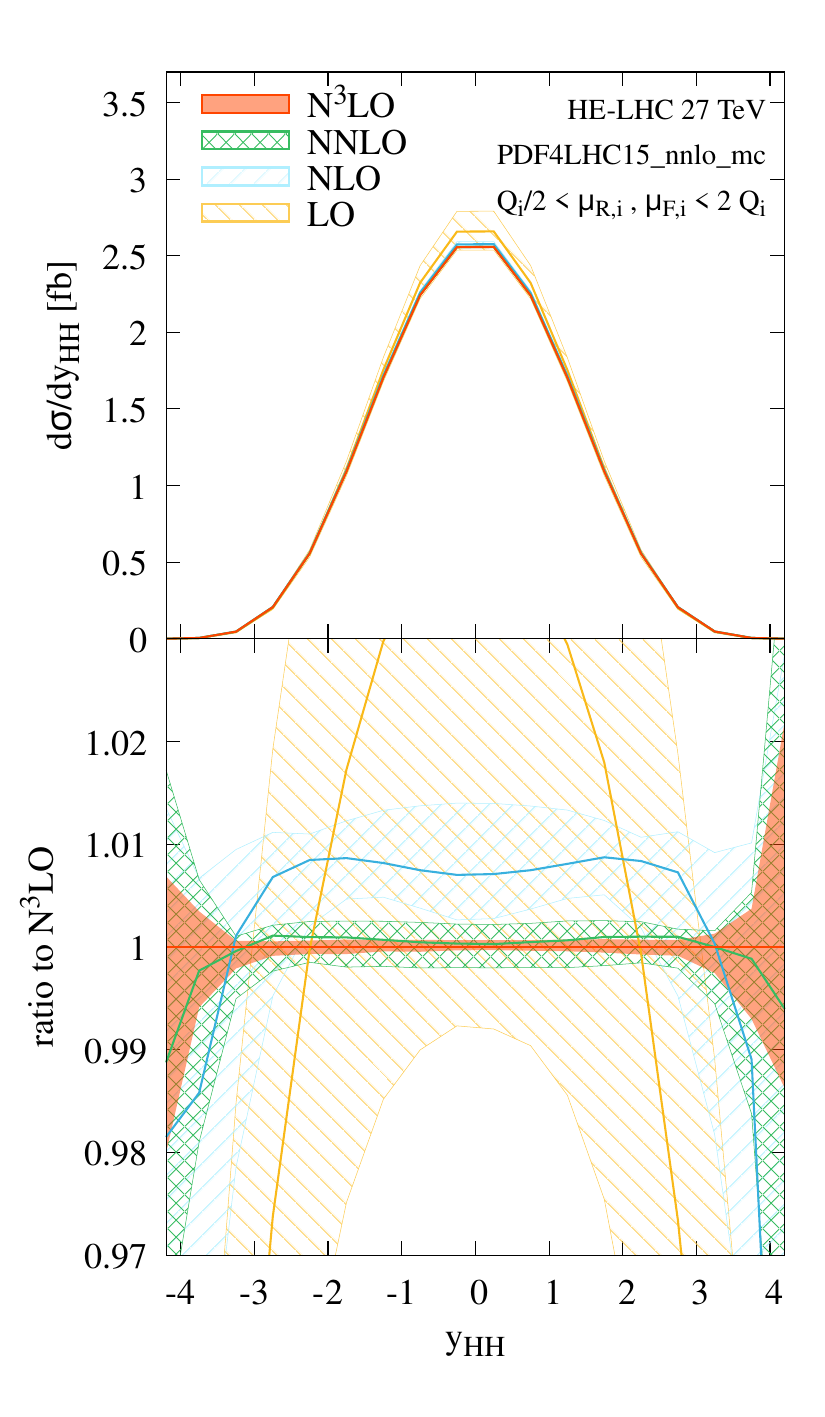}%
  \includegraphics[clip,width=0.33\textwidth,page=8,angle=0]{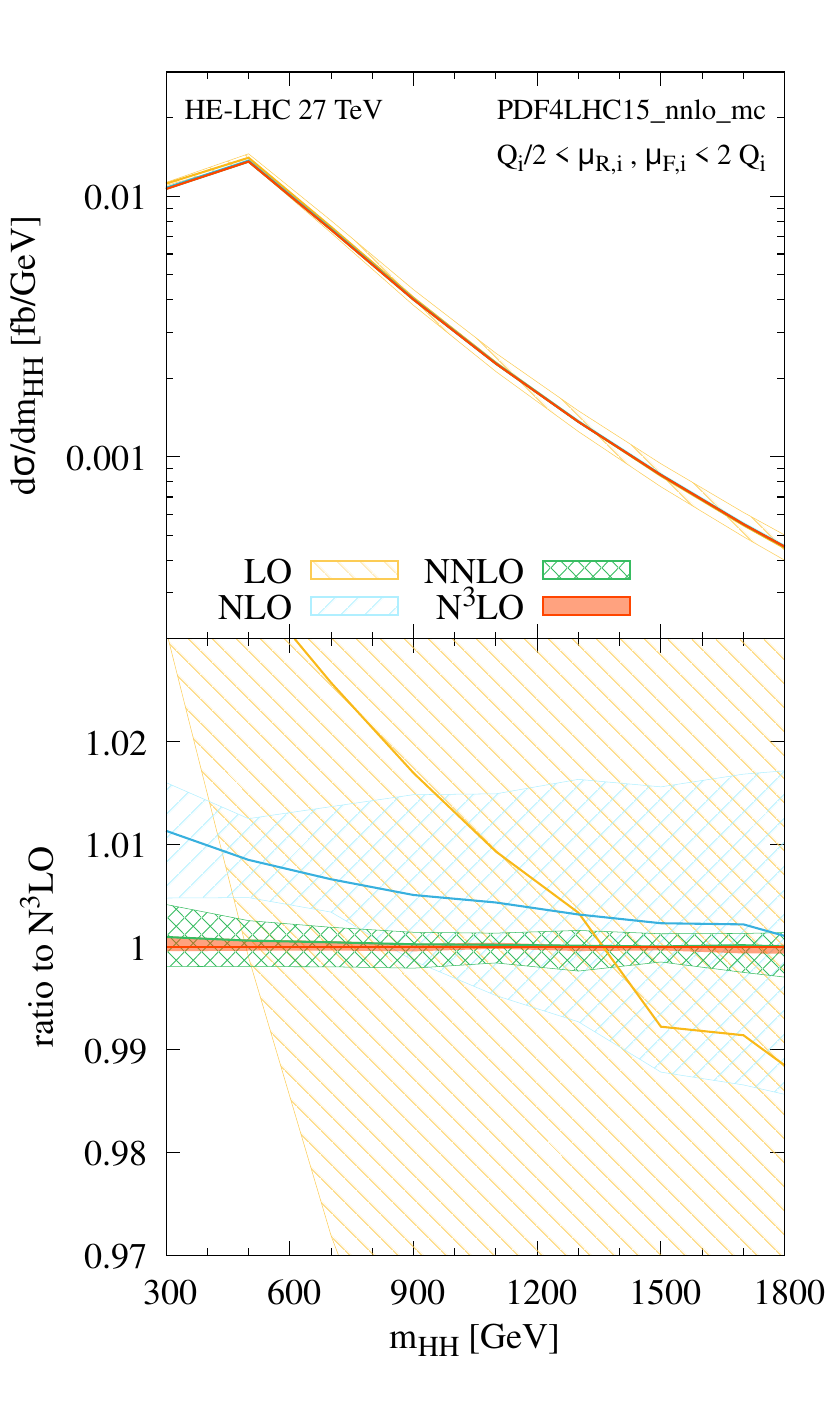}%
  \caption{Differential cross sections for the transverse momentum
    $p_{t,HH}$, rapidity $y_{HH}$ and mass $m_{HH}$ distributions of the Higgs
    pair.}
  \label{fig:diff-cross-sections-HH}
\end{figure*}

In particular, the $s$-channel production mode, while suppressed to a
few permille after VBF cuts, contributes to about $16\%$ to the total
cross section for 27 TeV collisions, and can therefore not be
neglected.
It can be calculated to NLO using the \texttt{MadGraph5\_aMC@NLO}
framework~\cite{Alwall:2014hca} and can be straightforwardly included.

Furthermore, NLO electroweak corrections are currently unknown and
expected to be sizeable.
They can be estimated from dominant light quark induced channels using
\texttt{Recola(Collier)+MoCaNLO}~\cite{Actis:2016mpe,Denner:2016kdg,MoCaNLO,Bendavid:2018nar}
for the di-Higgs and single-Higgs VBF process, comparing the
latter to \texttt{HAWK}~\cite{Denner:2014cla}.
For VBF Higgs pair production the EW corrections to the inclusive
cross section lie between $-5\%$ and $-7\%$.
Compared to the single-Higgs VBF correction of roughly $-5\%$ (using
the same set-up, i.e. excluding photonic and b-quark channels), the
double Higgs VBF process thus does not seem to receive large VBS-like
corrections~\cite{Biedermann:2016yds,Biedermann:2017bss}.
One can therefore expect the full electroweak corrections to be at
least at the same level as the NNLO QCD corrections, and significantly
larger than the \NNNLO{} corrections.

There are also a number of $\as^2$ and $\as^3$ contributions that are
neglected in the structure function approximation, notably: the double
and triple gluon exchange between the two quark lines; heavy-quark
loop induced production; $t$-/$u$-channel interferences; single-quark
line contributions; loop induced interferences between VBF and
gluon-fusion Higgs production.
These have been shown to contribute at the few permille level in
single Higgs VBF
production~\cite{Ciccolini:2007ec,Andersen:2007mp,Harlander:2008xn,Bolzoni:2011cu},
and we therefore expect that they can be neglected.

The impact on the cross section of PDF and $\as$ uncertainties can be
evaluated using the \texttt{PDF4LHC15\_nnlo\_mc\_pdfas} set, and is
of about $2.1\%$.
Finally, there is also a theoretical PDF uncertainty, due to missing
higher orders in the determination of the PDFs. 
In this paper we use an NNLO pdf set to evaluate an \NNNLO{} cross
section, since \NNNLO{} sets are currently unavailable. 
The uncertainty due to these missing higher order terms come from two
sources. They are dominated by missing third order corrections to the
coefficient functions relating physical observables to PDFs, and can
be estimated to about $8\permil$ using the method presented
in~\cite{Dreyer:2016oyx}. The second source of corrections is due to
unknown four-loop splitting functions~\cite{Vogt:2018miu} appearing in
the DGLAP evolution, which have been estimated to be
negligible~\cite{Dreyer:2016oyx}.

\section{Differential distributions}
\label{sec:diff-dist}

The calculation described in section~\ref{sec:hh} is inclusive over
final state QCD radiation.
One can thus not obtain differential predictions with respect to the
jet kinematics without using the projection-to-Born
method~\cite{Cacciari:2015jma} and combining it with a higher
multiplicity NNLO prediction.
However, we have full access to the kinematics of the Higgs bosons,
and it is therefore straightforward to compute differential
observables with respect to the their momenta.
Let us now focus on several differential distributions of particular
interest.
We will again consider here a 27 TeV proton-proton collider except
where otherwise specified.

\begin{figure*}
  \centering
  \includegraphics[clip,width=0.33\textwidth,page=3,angle=0]{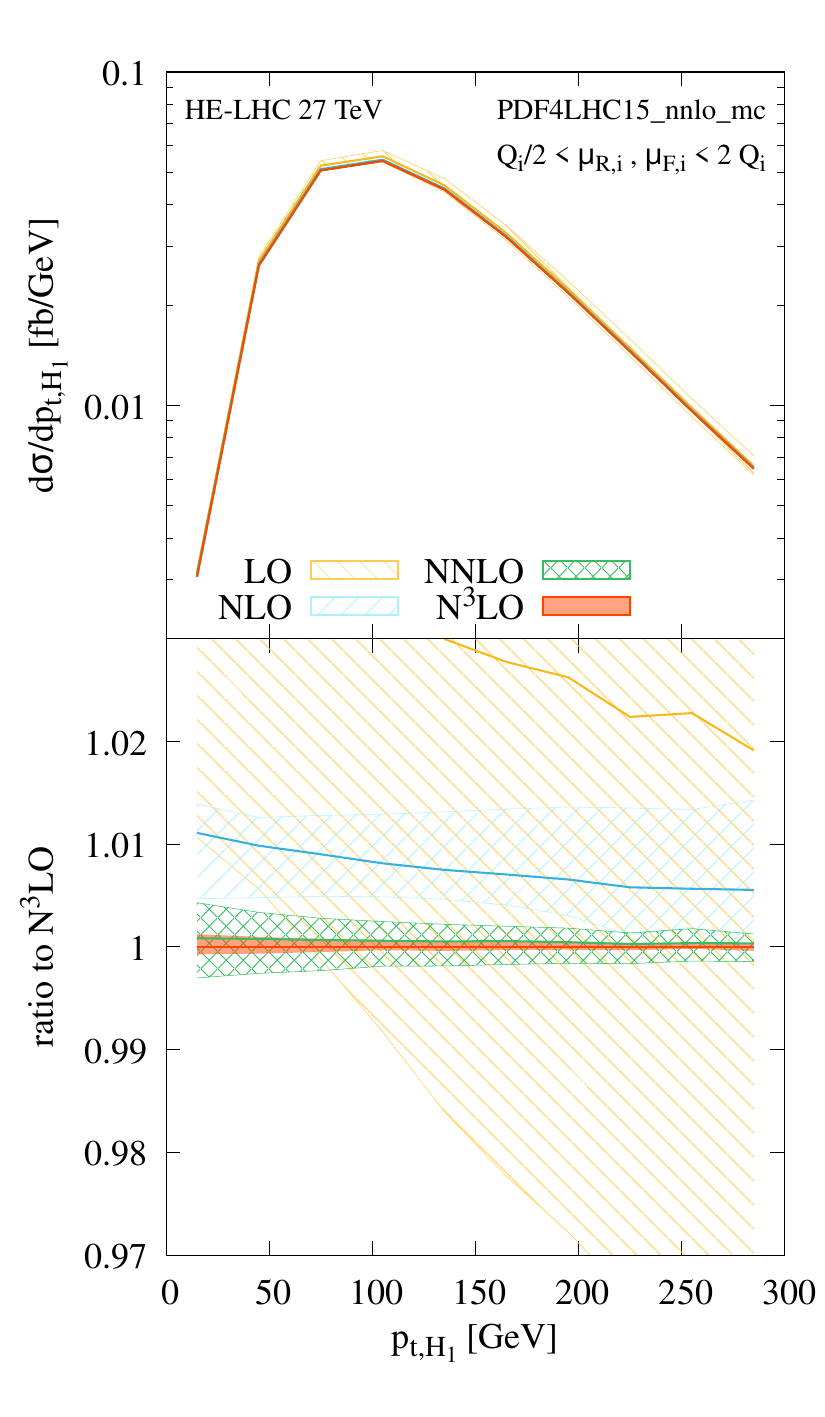}%
  \hspace{7mm}\includegraphics[clip,width=0.33\textwidth,page=4,angle=0]{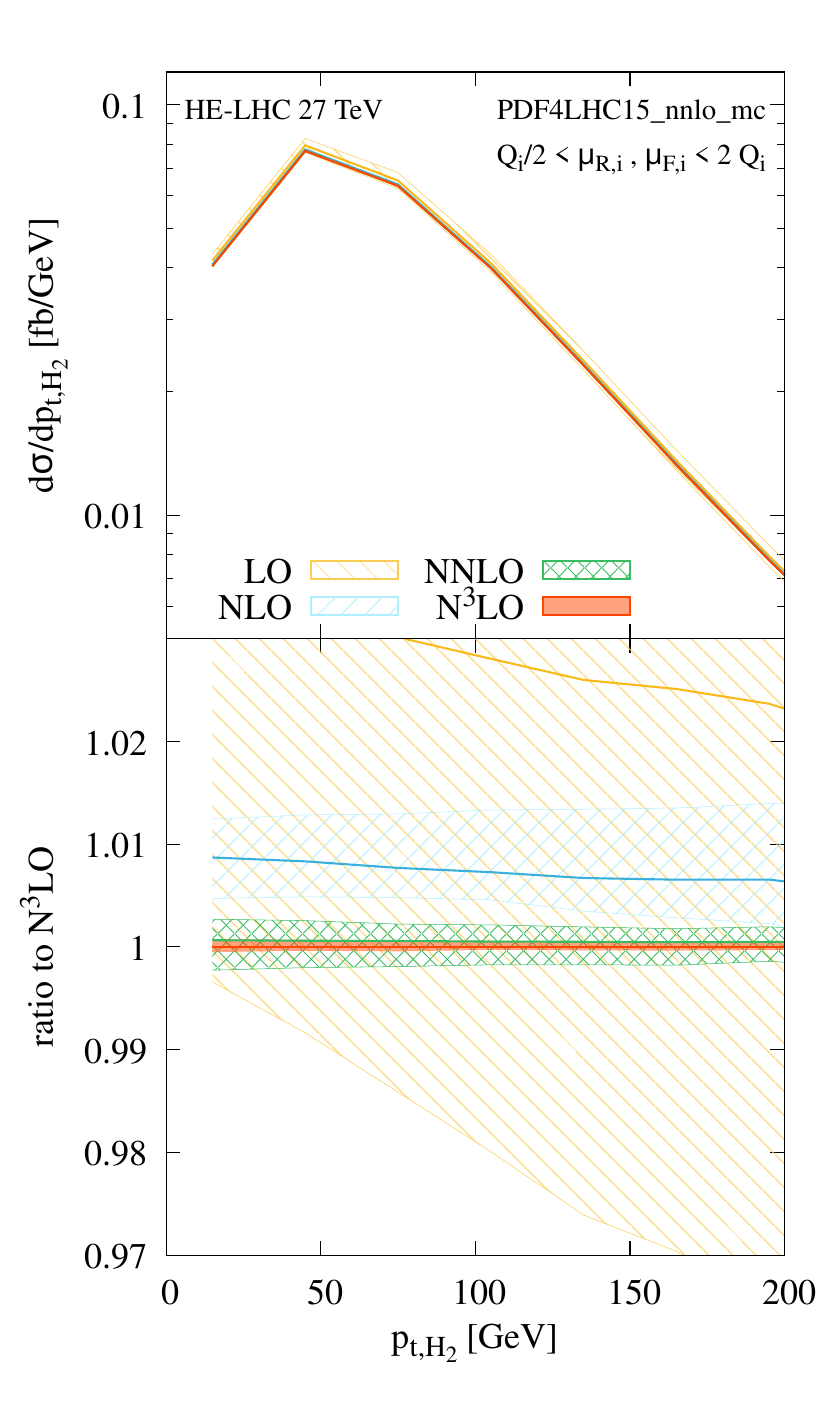}%
  \caption{Differential cross sections for the transverse momentum
    distributions for both the harder ($p_{t,H_1}$) and softer
    ($p_{t,H_2}$) Higgs boson.}
  \label{fig:diff-cross-sections-pt}
\end{figure*}

\begin{figure*}
  \centering
  \includegraphics[clip,width=0.33\textwidth,page=6,angle=0]{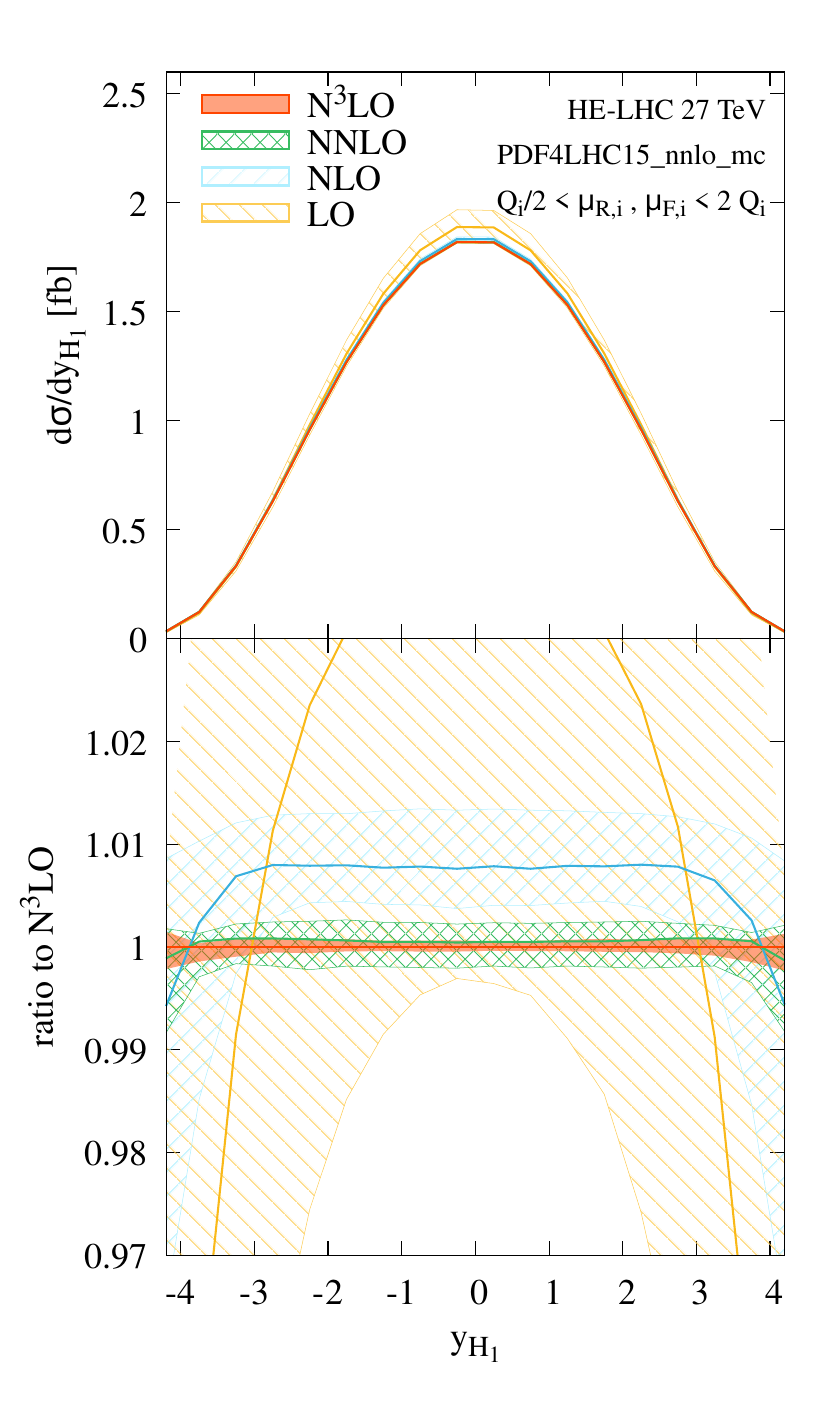}%
  \hspace{7mm}\includegraphics[clip,width=0.33\textwidth,page=7,angle=0]{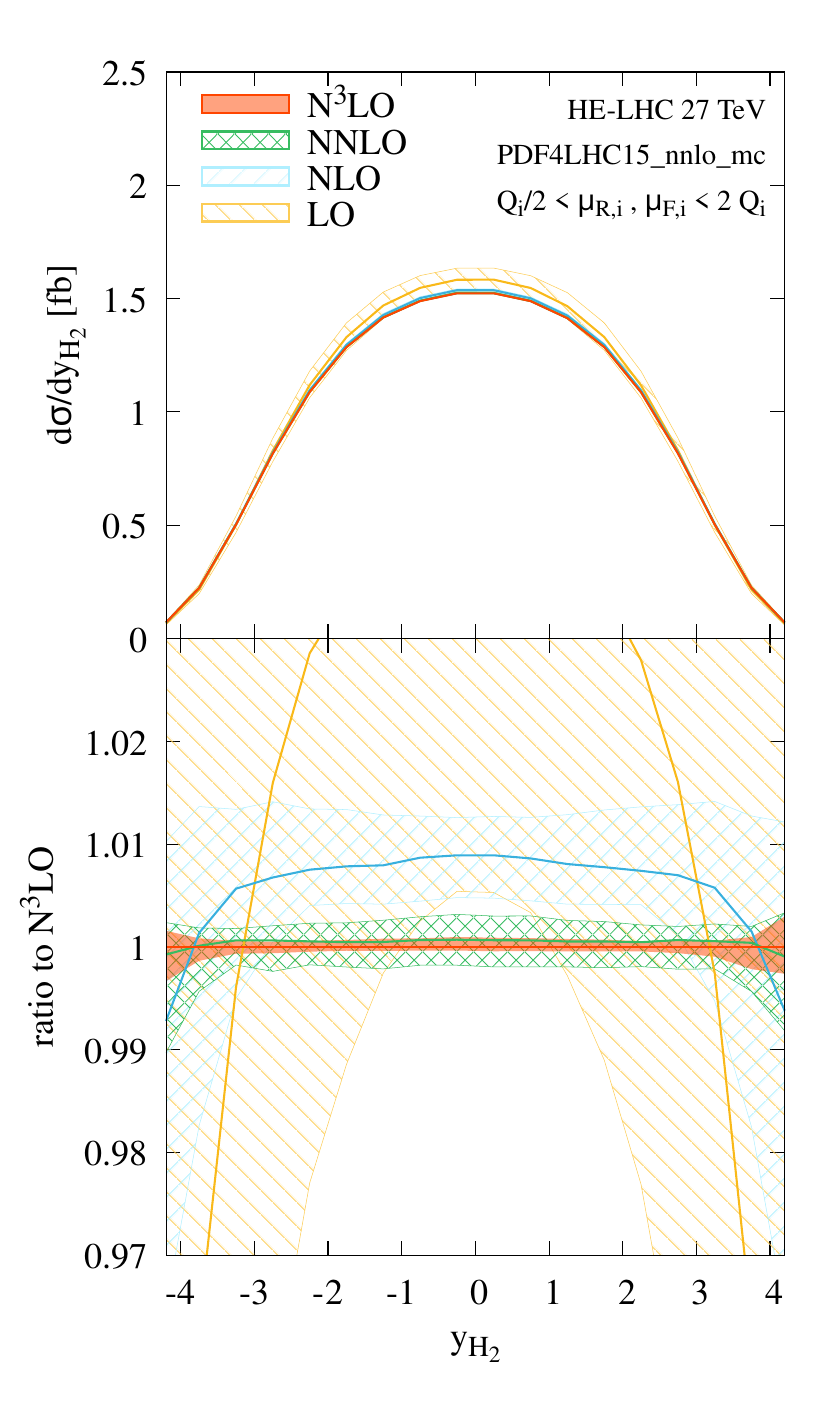}%
  \caption{Differential cross sections for the rapidity distributions for both the harder ($y_{H_1}$)
    and softer ($y_{H_2}$) Higgs boson.}
  \label{fig:diff-cross-sections-rap}
\end{figure*}

\begin{figure*}
  \centering
  \includegraphics[clip,width=0.45\textwidth,page=1,angle=0]{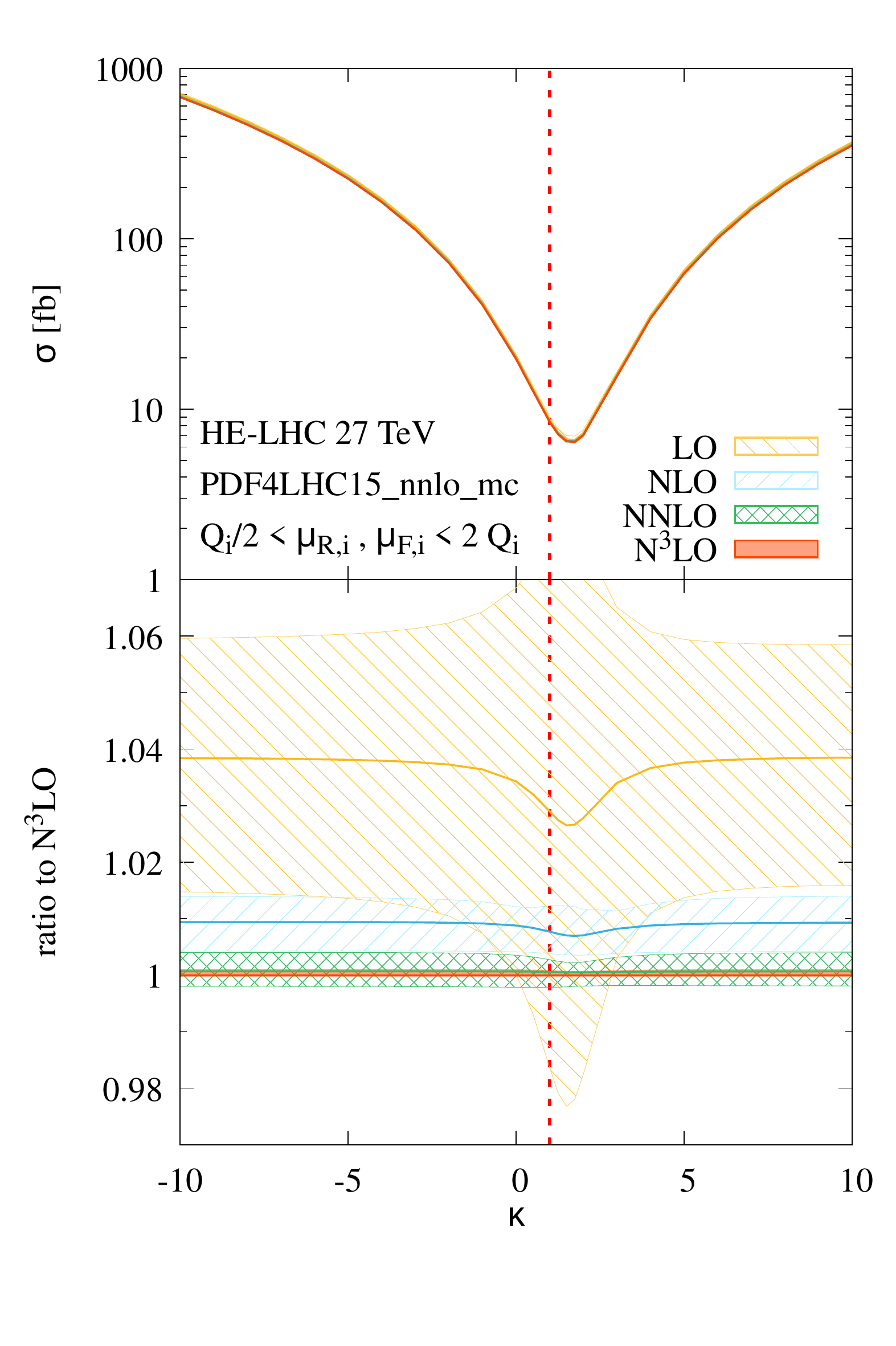}%
  \hspace{6mm}%
  \includegraphics[clip,width=0.45\textwidth,page=1,angle=0]{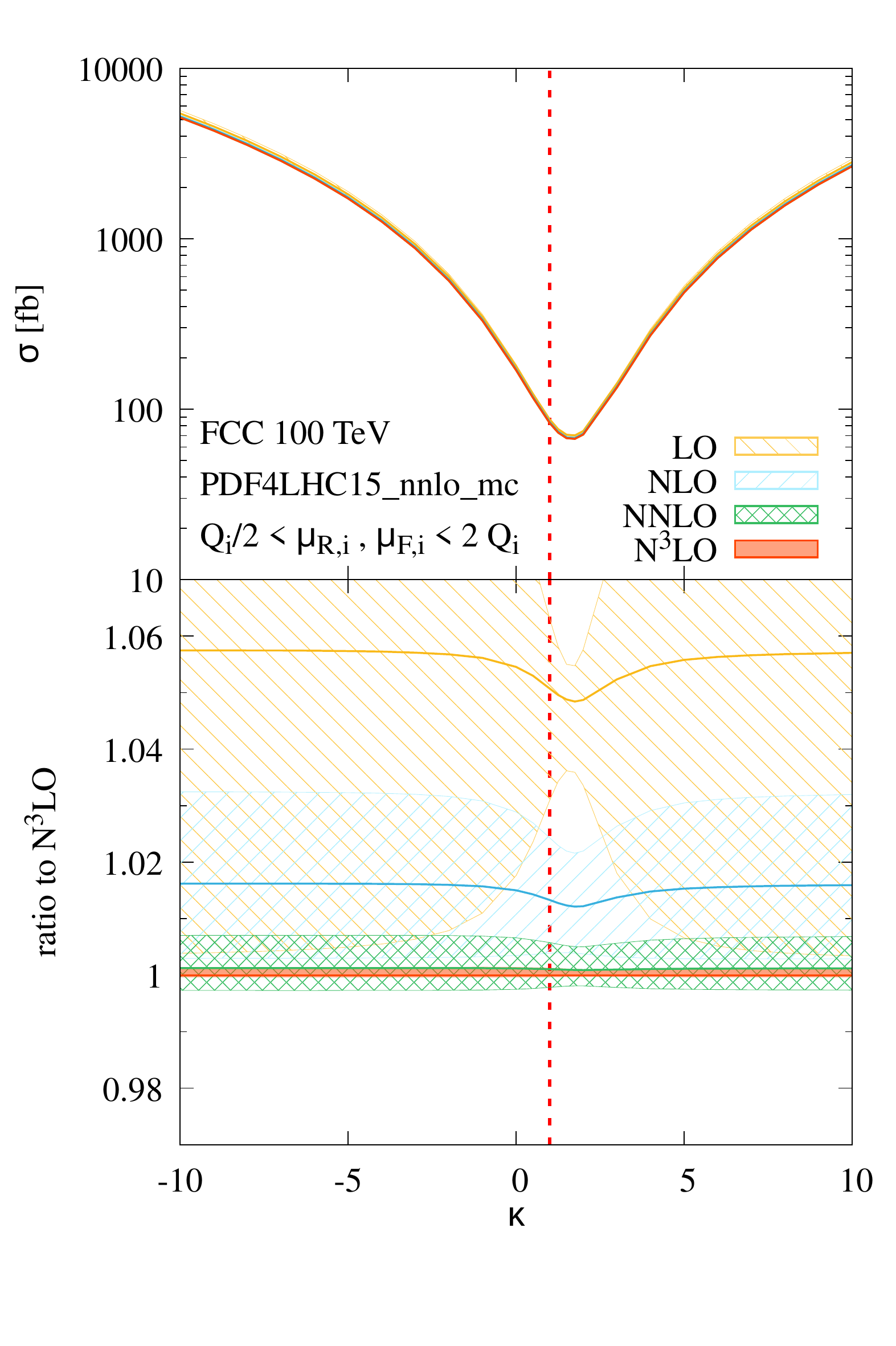}
  \caption{Total cross section as a function of $\kappa$ for both the 27 TeV HE-LHC
    (left) and the 100 TeV FCC (right).
    The lower panel gives the ratio to the central \NNNLO value, with
    scale uncertainty bands obtained from a seven-point variation.  }
  \label{fig:total-cross-sections-lambda}
\end{figure*}

In figure~\ref{fig:diff-cross-sections-HH}, we show the transverse
momentum $p_{t,HH}$, rapidity $y_{HH}$ and invariant mass $m_{HH}$
distributions of the Higgs pair for each order in QCD.
The latter is of particular interest, as the Higgs pair invariant mass
can be notably sensitive to deviations due to physics beyond the
SM~\cite{Kaplan:1983fs,Contino:2010mh,Bishara:2016kjn}.
We see that once we get to the third order, there is almost no
kinematic dependence to the $K$-factor, except at very high
rapidities, where the \NNNLO{} corrections can bring changes to the
central value of about one percent.
The \NNNLO{} scale variation bands are always fully contained within
the NNLO scale uncertainties, but are about four times thinner.

We order the Higgs bosons according to their transverse
momentum. Figure~\ref{fig:diff-cross-sections-pt} provides the
transverse momentum distribution of both the harder ($p_{t,H_1}$) and
the softer ($p_{t,H_2}$) Higgs.
The third order corrections to these observables are negligible,
with again a large reduction in scale uncertainties.
The corrections to the rapidity distributions of the two Higgs bosons
are shown in figure~\ref{fig:diff-cross-sections-rap}.
We note here that the $\as^3$ contribution has almost no kinematic
dependence up to rapidities of $|y_H|\sim4$, with the scale variation
bands being again fully contained by the theoretical uncertainties of
the previous order.

Finally, let us study the impact of the trilinear Higgs self coupling,
$\lambda$, by varying the corresponding factor in
equation~(\ref{eq:VV-subproc}).
Constraining the trilinear coupling is of particular interest, since
many scenarios of new physics beyond the SM predict significant
deviations of this value.
Examples of such models are $SO(5)/SO(4)$ minimal composite
Higgs~\cite{Agashe:2004rs,Contino:2006qr} and dilaton
models~\cite{Goldberger:2008zz}.

To study the impact of deviations of this type, we define
$\lambda = \kappa \lambda^\text{SM}$, where
$\lambda^\text{SM}=m_h^2 / 2v$, and consider a range of values
for $\kappa$.

The total cross section up to \NNNLO{} as a function of $\kappa$ is
given in figure~\ref{fig:total-cross-sections-lambda}, both for a 27
TeV HE-LHC and for a 100 TeV FCC.
One can observe that while the inclusive cross section changes by
several orders of magnitude, there is as expected almost no dependence
of the higher order corrections on $\kappa$ beyond leading order.

\begin{figure*}
  \centering
  \includegraphics[clip,width=0.33\textwidth,page=1,angle=0]{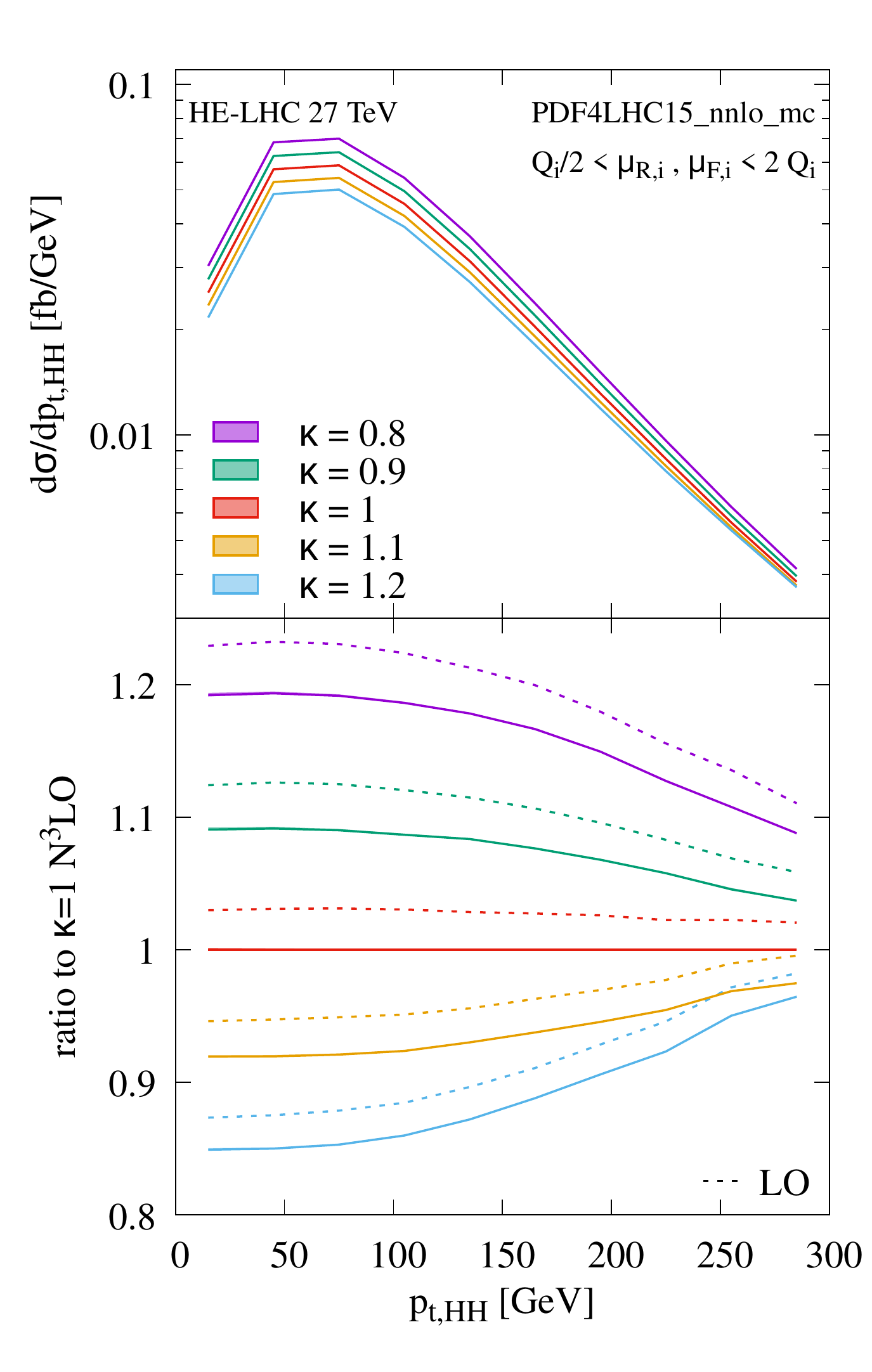}%
  \includegraphics[clip,width=0.33\textwidth,page=4,angle=0]{lambdavar.pdf}%
  \includegraphics[clip,width=0.33\textwidth,page=7,angle=0]{lambdavar.pdf}%
  \caption{Differential cross sections for the transverse momentum
    $p_{t,HH}$, rapidity $y_{HH}$ and mass $m_{HH}$ distributions of the Higgs pair
    at \NNNLO{} for different values of the trilinear Higgs coupling
    $\lambda$.
    The lower panel gives the ratio to the central \NNNLO{} prediction
    with $\lambda=\lambda^\text{SM}$, with the LO shown as dashed lines.
    The bands correspond to theoretical scale uncertainties.}
  \label{fig:lambda-cross-sections}
\end{figure*}

In figure~\ref{fig:lambda-cross-sections}, we show the kinematics of
the Higgs pair for several values of $\kappa$.
We see that even very small deviations in the trilinear Higgs coupling
have a substantial impact on the cross sections, both in the
normalization and shape of the distributions.
In particular, the rapidity and invariant mass of the Higgs pair are
particularly sensitive to changes in $\lambda$.

The lower panels in figure~\ref{fig:lambda-cross-sections} show the
ratio to the central value obtained with $\kappa=1$, with the leading
order predictions shown as dashed lines.
One can see that the changes to the cross sections from variations in
$\kappa$ can be substantial.
However there is essentially no change in the \NNNLO{}/LO $K$-factor.


\section{Conclusions}
\label{sec:conclusion}

In this article, we have completed the first \NNNLO{} calculation of a
$2\to4$ process, namely the production of a Higgs boson pair through VBF.
This calculation was made possible by the factorizable nature of the
higher order QCD corrections.
Together with the fully differential NNLO calculation presented in a
companion paper~\cite{Dreyer:2018rfu}, this brings the di-Higgs
production channel to the same theoretical accuracy as has been
achieved for the single-Higgs process, opening up the prospect of
precision studies of the Higgs sector through Higgs pair production at
the HE-LHC and at future hadron colliders.

We have presented differential distributions of the Higgs pair
transverse momentum, rapidity and invariant mass for the 27 TeV
HE-LHC.
The corrections are at the few permille level, however the calculation
of the third order leads to a substantial reduction in scale
uncertainties.
The convergence of the perturbative series is very stable at this
order, with almost no kinematic dependence to the \NNNLO{}
corrections, except at very high rapidities.
The \NNNLO{} scale variation bands are always fully contained within
the second order scale uncertainties, but are over four times thinner.

Finally we studied the impact of deviations from the SM in the
trilinear Higgs self-coupling on the \NNNLO{} distributions.
Small deviations of this constant can substantially change both
the total cross section and the shape of the distributions.
However the structure of the higher order QCD corrections is
unaffected by variations in the coupling, with the \NNNLO{}/LO
$K$-factor staying constant over a range broad range of $\lambda$
values.

The results presented here have been implemented in the version 2.0.0
of
\href{https://provbfh.hepforge.org/}{\texttt{proVBFH-incl}}~\cite{proVBFH}
which provides predictions for both single and double Higgs inclusive
cross sections up to \NNNLO{} in QCD.

This article provides also the first element for a fully differential
\NNNLO{} calculation of VBF Higgs pair production.
This could be achieved by combining the present inclusive calculation
with a differential NNLO computation of the electroweak production of
two Higgs bosons in association with three jets.

\textbf{Acknowledgments:}
We are grateful to Jean-Nicolas Lang and Mathieu Pellen for providing
us with an estimate of the electroweak corrections.
We also thank Gavin Salam and Giulia Zanderighi for their careful
reading of the manuscript and useful comments.
F.D.\ thanks the University of Zurich and the Pauli Center for
Theoretical Studies, and A.K. thanks the University of Oxford and the
Rudolf Peierls Centre for Theoretical Physics for hospitality while
this work was being completed.
F.D.\ is supported by the Science and Technology Facilities Council
(STFC) under grant ST/P000770/1.
A.K.\ is supported by the Swiss National Science Foundation (SNF)
under grant number 200020-175595.
%

\bibliography{dihiggs}

\begin{thebibliography}{60}
\expandafter\ifx\csname natexlab\endcsname\relax\def\natexlab#1{#1}\fi
\expandafter\ifx\csname bibnamefont\endcsname\relax
  \def\bibnamefont#1{#1}\fi
\expandafter\ifx\csname bibfnamefont\endcsname\relax
  \def\bibfnamefont#1{#1}\fi
\expandafter\ifx\csname citenamefont\endcsname\relax
  \def\citenamefont#1{#1}\fi
\expandafter\ifx\csname url\endcsname\relax
  \def\url#1{\texttt{#1}}\fi
\expandafter\ifx\csname urlprefix\endcsname\relax\def\urlprefix{URL }\fi
\providecommand{\bibinfo}[2]{#2}
\providecommand{\eprint}[2][]{\url{#2}}

\bibitem[{\citenamefont{Aad et~al.}(2012)}]{Aad:2012tfa}
\bibinfo{author}{\bibfnamefont{G.}~\bibnamefont{Aad}} \bibnamefont{et~al.}
  (\bibinfo{collaboration}{ATLAS}), \bibinfo{journal}{Phys. Lett.}
  \textbf{\bibinfo{volume}{B716}}, \bibinfo{pages}{1} (\bibinfo{year}{2012}),
  \eprint{1207.7214}.

\bibitem[{\citenamefont{Chatrchyan et~al.}(2012)}]{Chatrchyan:2012xdj}
\bibinfo{author}{\bibfnamefont{S.}~\bibnamefont{Chatrchyan}}
  \bibnamefont{et~al.} (\bibinfo{collaboration}{CMS}), \bibinfo{journal}{Phys.
  Lett.} \textbf{\bibinfo{volume}{B716}}, \bibinfo{pages}{30}
  (\bibinfo{year}{2012}), \eprint{1207.7235}.

\bibitem[{\citenamefont{Aaboud et~al.}(2018{\natexlab{a}})}]{Aaboud:2018sfw}
\bibinfo{author}{\bibfnamefont{M.}~\bibnamefont{Aaboud}} \bibnamefont{et~al.}
  (\bibinfo{collaboration}{ATLAS}), \bibinfo{journal}{Submitted to: Phys. Rev.
  Lett.}  (\bibinfo{year}{2018}{\natexlab{a}}), \eprint{1808.00336}.

\bibitem[{\citenamefont{Aaboud et~al.}(2018{\natexlab{b}})}]{Aaboud:2018ewm}
\bibinfo{author}{\bibfnamefont{M.}~\bibnamefont{Aaboud}} \bibnamefont{et~al.}
  (\bibinfo{collaboration}{ATLAS}), \bibinfo{journal}{Submitted to: Eur. Phys.
  J.}  (\bibinfo{year}{2018}{\natexlab{b}}), \eprint{1807.08567}.

\bibitem[{\citenamefont{Aaboud et~al.}(2018{\natexlab{c}})}]{Aaboud:2018ftw}
\bibinfo{author}{\bibfnamefont{M.}~\bibnamefont{Aaboud}} \bibnamefont{et~al.}
  (\bibinfo{collaboration}{ATLAS}) (\bibinfo{year}{2018}{\natexlab{c}}),
  \eprint{1807.04873}.

\bibitem[{\citenamefont{Aaboud et~al.}(2018{\natexlab{d}})}]{Aaboud:2018knk}
\bibinfo{author}{\bibfnamefont{M.}~\bibnamefont{Aaboud}} \bibnamefont{et~al.}
  (\bibinfo{collaboration}{ATLAS}) (\bibinfo{year}{2018}{\natexlab{d}}),
  \eprint{1804.06174}.

\bibitem[{\citenamefont{Aaboud et~al.}(2016)}]{Aaboud:2016xco}
\bibinfo{author}{\bibfnamefont{M.}~\bibnamefont{Aaboud}} \bibnamefont{et~al.}
  (\bibinfo{collaboration}{ATLAS}), \bibinfo{journal}{Phys. Rev.}
  \textbf{\bibinfo{volume}{D94}}, \bibinfo{pages}{052002}
  (\bibinfo{year}{2016}), \eprint{1606.04782}.

\bibitem[{\citenamefont{Aad et~al.}(2015{\natexlab{a}})}]{Aad:2015xja}
\bibinfo{author}{\bibfnamefont{G.}~\bibnamefont{Aad}} \bibnamefont{et~al.}
  (\bibinfo{collaboration}{ATLAS}), \bibinfo{journal}{Phys. Rev.}
  \textbf{\bibinfo{volume}{D92}}, \bibinfo{pages}{092004}
  (\bibinfo{year}{2015}{\natexlab{a}}), \eprint{1509.04670}.

\bibitem[{\citenamefont{Aad et~al.}(2015{\natexlab{b}})}]{Aad:2015uka}
\bibinfo{author}{\bibfnamefont{G.}~\bibnamefont{Aad}} \bibnamefont{et~al.}
  (\bibinfo{collaboration}{ATLAS}), \bibinfo{journal}{Eur. Phys. J.}
  \textbf{\bibinfo{volume}{C75}}, \bibinfo{pages}{412}
  (\bibinfo{year}{2015}{\natexlab{b}}), \eprint{1506.00285}.

\bibitem[{\citenamefont{Aad et~al.}(2015{\natexlab{c}})}]{Aad:2014yja}
\bibinfo{author}{\bibfnamefont{G.}~\bibnamefont{Aad}} \bibnamefont{et~al.}
  (\bibinfo{collaboration}{ATLAS}), \bibinfo{journal}{Phys. Rev. Lett.}
  \textbf{\bibinfo{volume}{114}}, \bibinfo{pages}{081802}
  (\bibinfo{year}{2015}{\natexlab{c}}), \eprint{1406.5053}.

\bibitem[{\citenamefont{Sirunyan
  et~al.}(2018{\natexlab{a}})}]{Sirunyan:2018tki}
\bibinfo{author}{\bibfnamefont{A.~M.} \bibnamefont{Sirunyan}}
  \bibnamefont{et~al.} (\bibinfo{collaboration}{CMS})
  (\bibinfo{year}{2018}{\natexlab{a}}), \eprint{1810.11854}.

\bibitem[{\citenamefont{Sirunyan
  et~al.}(2018{\natexlab{b}})}]{Sirunyan:2018iwt}
\bibinfo{author}{\bibfnamefont{A.~M.} \bibnamefont{Sirunyan}}
  \bibnamefont{et~al.} (\bibinfo{collaboration}{CMS})
  (\bibinfo{year}{2018}{\natexlab{b}}), \eprint{1806.00408}.

\bibitem[{\citenamefont{Sirunyan
  et~al.}(2018{\natexlab{c}})}]{Sirunyan:2017guj}
\bibinfo{author}{\bibfnamefont{A.~M.} \bibnamefont{Sirunyan}}
  \bibnamefont{et~al.} (\bibinfo{collaboration}{CMS}), \bibinfo{journal}{JHEP}
  \textbf{\bibinfo{volume}{01}}, \bibinfo{pages}{054}
  (\bibinfo{year}{2018}{\natexlab{c}}), \eprint{1708.04188}.

\bibitem[{\citenamefont{Sirunyan
  et~al.}(2018{\natexlab{d}})}]{Sirunyan:2017djm}
\bibinfo{author}{\bibfnamefont{A.~M.} \bibnamefont{Sirunyan}}
  \bibnamefont{et~al.} (\bibinfo{collaboration}{CMS}), \bibinfo{journal}{Phys.
  Lett.} \textbf{\bibinfo{volume}{B778}}, \bibinfo{pages}{101}
  (\bibinfo{year}{2018}{\natexlab{d}}), \eprint{1707.02909}.

\bibitem[{\citenamefont{Sirunyan et~al.}(2017)}]{Sirunyan:2017tqo}
\bibinfo{author}{\bibfnamefont{A.~M.} \bibnamefont{Sirunyan}}
  \bibnamefont{et~al.} (\bibinfo{collaboration}{CMS}), \bibinfo{journal}{Phys.
  Rev.} \textbf{\bibinfo{volume}{D96}}, \bibinfo{pages}{072004}
  (\bibinfo{year}{2017}), \eprint{1707.00350}.

\bibitem[{\citenamefont{de~Florian
  et~al.}(2016{\natexlab{a}})}]{deFlorian:2016spz}
\bibinfo{author}{\bibfnamefont{D.}~\bibnamefont{de~Florian}}
  \bibnamefont{et~al.} (\bibinfo{collaboration}{LHC Higgs Cross Section Working
  Group}) (\bibinfo{year}{2016}{\natexlab{a}}), \eprint{1610.07922}.

\bibitem[{\citenamefont{de~Florian and Mazzitelli}(2013)}]{deFlorian:2013jea}
\bibinfo{author}{\bibfnamefont{D.}~\bibnamefont{de~Florian}} \bibnamefont{and}
  \bibinfo{author}{\bibfnamefont{J.}~\bibnamefont{Mazzitelli}},
  \bibinfo{journal}{Phys. Rev. Lett.} \textbf{\bibinfo{volume}{111}},
  \bibinfo{pages}{201801} (\bibinfo{year}{2013}), \eprint{1309.6594}.

\bibitem[{\citenamefont{de~Florian
  et~al.}(2016{\natexlab{b}})\citenamefont{de~Florian, Grazzini, Hanga,
  Kallweit, Lindert, Maierhöfer, Mazzitelli, and Rathlev}}]{deFlorian:2016uhr}
\bibinfo{author}{\bibfnamefont{D.}~\bibnamefont{de~Florian}},
  \bibinfo{author}{\bibfnamefont{M.}~\bibnamefont{Grazzini}},
  \bibinfo{author}{\bibfnamefont{C.}~\bibnamefont{Hanga}},
  \bibinfo{author}{\bibfnamefont{S.}~\bibnamefont{Kallweit}},
  \bibinfo{author}{\bibfnamefont{J.~M.} \bibnamefont{Lindert}},
  \bibinfo{author}{\bibfnamefont{P.}~\bibnamefont{Maierhöfer}},
  \bibinfo{author}{\bibfnamefont{J.}~\bibnamefont{Mazzitelli}},
  \bibnamefont{and} \bibinfo{author}{\bibfnamefont{D.}~\bibnamefont{Rathlev}},
  \bibinfo{journal}{JHEP} \textbf{\bibinfo{volume}{09}}, \bibinfo{pages}{151}
  (\bibinfo{year}{2016}{\natexlab{b}}), \eprint{1606.09519}.

\bibitem[{\citenamefont{de~Florian and Mazzitelli}(2015)}]{deFlorian:2015moa}
\bibinfo{author}{\bibfnamefont{D.}~\bibnamefont{de~Florian}} \bibnamefont{and}
  \bibinfo{author}{\bibfnamefont{J.}~\bibnamefont{Mazzitelli}},
  \bibinfo{journal}{JHEP} \textbf{\bibinfo{volume}{09}}, \bibinfo{pages}{053}
  (\bibinfo{year}{2015}), \eprint{1505.07122}.

\bibitem[{\citenamefont{Grazzini et~al.}(2018)\citenamefont{Grazzini, Heinrich,
  Jones, Kallweit, Kerner, Lindert, and Mazzitelli}}]{Grazzini:2018bsd}
\bibinfo{author}{\bibfnamefont{M.}~\bibnamefont{Grazzini}},
  \bibinfo{author}{\bibfnamefont{G.}~\bibnamefont{Heinrich}},
  \bibinfo{author}{\bibfnamefont{S.}~\bibnamefont{Jones}},
  \bibinfo{author}{\bibfnamefont{S.}~\bibnamefont{Kallweit}},
  \bibinfo{author}{\bibfnamefont{M.}~\bibnamefont{Kerner}},
  \bibinfo{author}{\bibfnamefont{J.~M.} \bibnamefont{Lindert}},
  \bibnamefont{and}
  \bibinfo{author}{\bibfnamefont{J.}~\bibnamefont{Mazzitelli}},
  \bibinfo{journal}{JHEP} \textbf{\bibinfo{volume}{05}}, \bibinfo{pages}{059}
  (\bibinfo{year}{2018}), \eprint{1803.02463}.

\bibitem[{\citenamefont{Baglio et~al.}(2013)\citenamefont{Baglio, Djouadi,
  Gröber, Mühlleitner, Quevillon, and Spira}}]{Baglio:2012np}
\bibinfo{author}{\bibfnamefont{J.}~\bibnamefont{Baglio}},
  \bibinfo{author}{\bibfnamefont{A.}~\bibnamefont{Djouadi}},
  \bibinfo{author}{\bibfnamefont{R.}~\bibnamefont{Gröber}},
  \bibinfo{author}{\bibfnamefont{M.~M.} \bibnamefont{Mühlleitner}},
  \bibinfo{author}{\bibfnamefont{J.}~\bibnamefont{Quevillon}},
  \bibnamefont{and} \bibinfo{author}{\bibfnamefont{M.}~\bibnamefont{Spira}},
  \bibinfo{journal}{JHEP} \textbf{\bibinfo{volume}{04}}, \bibinfo{pages}{151}
  (\bibinfo{year}{2013}), \eprint{1212.5581}.

\bibitem[{\citenamefont{Bishara et~al.}(2017)\citenamefont{Bishara, Contino,
  and Rojo}}]{Bishara:2016kjn}
\bibinfo{author}{\bibfnamefont{F.}~\bibnamefont{Bishara}},
  \bibinfo{author}{\bibfnamefont{R.}~\bibnamefont{Contino}}, \bibnamefont{and}
  \bibinfo{author}{\bibfnamefont{J.}~\bibnamefont{Rojo}},
  \bibinfo{journal}{Eur. Phys. J.} \textbf{\bibinfo{volume}{C77}},
  \bibinfo{pages}{481} (\bibinfo{year}{2017}), \eprint{1611.03860}.

\bibitem[{\citenamefont{Figy}(2008)}]{Figy:2008zd}
\bibinfo{author}{\bibfnamefont{T.}~\bibnamefont{Figy}}, \bibinfo{journal}{Mod.
  Phys. Lett.} \textbf{\bibinfo{volume}{A23}}, \bibinfo{pages}{1961}
  (\bibinfo{year}{2008}), \eprint{0806.2200}.

\bibitem[{\citenamefont{Frederix et~al.}(2014)\citenamefont{Frederix, Frixione,
  Hirschi, Maltoni, Mattelaer, Torrielli, Vryonidou, and
  Zaro}}]{Frederix:2014hta}
\bibinfo{author}{\bibfnamefont{R.}~\bibnamefont{Frederix}},
  \bibinfo{author}{\bibfnamefont{S.}~\bibnamefont{Frixione}},
  \bibinfo{author}{\bibfnamefont{V.}~\bibnamefont{Hirschi}},
  \bibinfo{author}{\bibfnamefont{F.}~\bibnamefont{Maltoni}},
  \bibinfo{author}{\bibfnamefont{O.}~\bibnamefont{Mattelaer}},
  \bibinfo{author}{\bibfnamefont{P.}~\bibnamefont{Torrielli}},
  \bibinfo{author}{\bibfnamefont{E.}~\bibnamefont{Vryonidou}},
  \bibnamefont{and} \bibinfo{author}{\bibfnamefont{M.}~\bibnamefont{Zaro}},
  \bibinfo{journal}{Phys. Lett.} \textbf{\bibinfo{volume}{B732}},
  \bibinfo{pages}{142} (\bibinfo{year}{2014}), \eprint{1401.7340}.

\bibitem[{\citenamefont{Ling et~al.}(2014)\citenamefont{Ling, Zhang, Ma, Guo,
  Li, and Li}}]{Liu-Sheng:2014gxa}
\bibinfo{author}{\bibfnamefont{L.-S.} \bibnamefont{Ling}},
  \bibinfo{author}{\bibfnamefont{R.-Y.} \bibnamefont{Zhang}},
  \bibinfo{author}{\bibfnamefont{W.-G.} \bibnamefont{Ma}},
  \bibinfo{author}{\bibfnamefont{L.}~\bibnamefont{Guo}},
  \bibinfo{author}{\bibfnamefont{W.-H.} \bibnamefont{Li}}, \bibnamefont{and}
  \bibinfo{author}{\bibfnamefont{X.-Z.} \bibnamefont{Li}},
  \bibinfo{journal}{Phys. Rev.} \textbf{\bibinfo{volume}{D89}},
  \bibinfo{pages}{073001} (\bibinfo{year}{2014}), \eprint{1401.7754}.

\bibitem[{\citenamefont{Dreyer and Karlberg}(2018)}]{Dreyer:2018rfu}
\bibinfo{author}{\bibfnamefont{F.~A.} \bibnamefont{Dreyer}} \bibnamefont{and}
  \bibinfo{author}{\bibfnamefont{A.}~\bibnamefont{Karlberg}}
  (\bibinfo{year}{2018}), \eprint{1811.07918}.

\bibitem[{\citenamefont{Bolzoni et~al.}(2010)\citenamefont{Bolzoni, Maltoni,
  Moch, and Zaro}}]{Bolzoni:2010xr}
\bibinfo{author}{\bibfnamefont{P.}~\bibnamefont{Bolzoni}},
  \bibinfo{author}{\bibfnamefont{F.}~\bibnamefont{Maltoni}},
  \bibinfo{author}{\bibfnamefont{S.-O.} \bibnamefont{Moch}}, \bibnamefont{and}
  \bibinfo{author}{\bibfnamefont{M.}~\bibnamefont{Zaro}},
  \bibinfo{journal}{Phys. Rev. Lett.} \textbf{\bibinfo{volume}{105}},
  \bibinfo{pages}{011801} (\bibinfo{year}{2010}), \eprint{1003.4451}.

\bibitem[{\citenamefont{Cacciari et~al.}(2015)\citenamefont{Cacciari, Dreyer,
  Karlberg, Salam, and Zanderighi}}]{Cacciari:2015jma}
\bibinfo{author}{\bibfnamefont{M.}~\bibnamefont{Cacciari}},
  \bibinfo{author}{\bibfnamefont{F.~A.} \bibnamefont{Dreyer}},
  \bibinfo{author}{\bibfnamefont{A.}~\bibnamefont{Karlberg}},
  \bibinfo{author}{\bibfnamefont{G.~P.} \bibnamefont{Salam}}, \bibnamefont{and}
  \bibinfo{author}{\bibfnamefont{G.}~\bibnamefont{Zanderighi}},
  \bibinfo{journal}{Phys. Rev. Lett.} \textbf{\bibinfo{volume}{115}},
  \bibinfo{pages}{082002} (\bibinfo{year}{2015}), \bibinfo{note}{[Erratum:
  Phys. Rev. Lett.120,no.13,139901(2018)]}, \eprint{1506.02660}.

\bibitem[{\citenamefont{Dreyer and Karlberg}(2016)}]{Dreyer:2016oyx}
\bibinfo{author}{\bibfnamefont{F.~A.} \bibnamefont{Dreyer}} \bibnamefont{and}
  \bibinfo{author}{\bibfnamefont{A.}~\bibnamefont{Karlberg}},
  \bibinfo{journal}{Phys. Rev. Lett.} \textbf{\bibinfo{volume}{117}},
  \bibinfo{pages}{072001} (\bibinfo{year}{2016}), \eprint{1606.00840}.

\bibitem[{\citenamefont{Cruz-Martinez et~al.}(2018)\citenamefont{Cruz-Martinez,
  Gehrmann, Glover, and Huss}}]{Cruz-Martinez:2018rod}
\bibinfo{author}{\bibfnamefont{J.}~\bibnamefont{Cruz-Martinez}},
  \bibinfo{author}{\bibfnamefont{T.}~\bibnamefont{Gehrmann}},
  \bibinfo{author}{\bibfnamefont{E.~W.~N.} \bibnamefont{Glover}},
  \bibnamefont{and} \bibinfo{author}{\bibfnamefont{A.}~\bibnamefont{Huss}},
  \bibinfo{journal}{Phys. Lett.} \textbf{\bibinfo{volume}{B781}},
  \bibinfo{pages}{672} (\bibinfo{year}{2018}), \eprint{1802.02445}.

\bibitem[{\citenamefont{Han et~al.}(1992)\citenamefont{Han, Valencia, and
  Willenbrock}}]{Han:1992hr}
\bibinfo{author}{\bibfnamefont{T.}~\bibnamefont{Han}},
  \bibinfo{author}{\bibfnamefont{G.}~\bibnamefont{Valencia}}, \bibnamefont{and}
  \bibinfo{author}{\bibfnamefont{S.}~\bibnamefont{Willenbrock}},
  \bibinfo{journal}{Phys. Rev. Lett.} \textbf{\bibinfo{volume}{69}},
  \bibinfo{pages}{3274} (\bibinfo{year}{1992}), \eprint{hep-ph/9206246}.

\bibitem[{\citenamefont{Ciccolini et~al.}(2008)\citenamefont{Ciccolini, Denner,
  and Dittmaier}}]{Ciccolini:2007ec}
\bibinfo{author}{\bibfnamefont{M.}~\bibnamefont{Ciccolini}},
  \bibinfo{author}{\bibfnamefont{A.}~\bibnamefont{Denner}}, \bibnamefont{and}
  \bibinfo{author}{\bibfnamefont{S.}~\bibnamefont{Dittmaier}},
  \bibinfo{journal}{Phys. Rev.} \textbf{\bibinfo{volume}{D77}},
  \bibinfo{pages}{013002} (\bibinfo{year}{2008}), \eprint{0710.4749}.

\bibitem[{\citenamefont{Harlander et~al.}(2008)\citenamefont{Harlander,
  Vollinga, and Weber}}]{Harlander:2008xn}
\bibinfo{author}{\bibfnamefont{R.~V.} \bibnamefont{Harlander}},
  \bibinfo{author}{\bibfnamefont{J.}~\bibnamefont{Vollinga}}, \bibnamefont{and}
  \bibinfo{author}{\bibfnamefont{M.~M.} \bibnamefont{Weber}},
  \bibinfo{journal}{Phys. Rev.} \textbf{\bibinfo{volume}{D77}},
  \bibinfo{pages}{053010} (\bibinfo{year}{2008}), \eprint{0801.3355}.

\bibitem[{\citenamefont{Bolzoni et~al.}(2012)\citenamefont{Bolzoni, Maltoni,
  Moch, and Zaro}}]{Bolzoni:2011cu}
\bibinfo{author}{\bibfnamefont{P.}~\bibnamefont{Bolzoni}},
  \bibinfo{author}{\bibfnamefont{F.}~\bibnamefont{Maltoni}},
  \bibinfo{author}{\bibfnamefont{S.-O.} \bibnamefont{Moch}}, \bibnamefont{and}
  \bibinfo{author}{\bibfnamefont{M.}~\bibnamefont{Zaro}},
  \bibinfo{journal}{Phys. Rev.} \textbf{\bibinfo{volume}{D85}},
  \bibinfo{pages}{035002} (\bibinfo{year}{2012}), \eprint{1109.3717}.

\bibitem[{\citenamefont{Dobrovolskaya and
  Novikov}(1991)}]{Dobrovolskaya:1990kx}
\bibinfo{author}{\bibfnamefont{A.}~\bibnamefont{Dobrovolskaya}}
  \bibnamefont{and} \bibinfo{author}{\bibfnamefont{V.}~\bibnamefont{Novikov}},
  \bibinfo{journal}{Z. Phys.} \textbf{\bibinfo{volume}{C52}},
  \bibinfo{pages}{427} (\bibinfo{year}{1991}).

\bibitem[{\citenamefont{Moch et~al.}(2005)\citenamefont{Moch, Vermaseren, and
  Vogt}}]{Moch:2004xu}
\bibinfo{author}{\bibfnamefont{S.}~\bibnamefont{Moch}},
  \bibinfo{author}{\bibfnamefont{J.~A.~M.} \bibnamefont{Vermaseren}},
  \bibnamefont{and} \bibinfo{author}{\bibfnamefont{A.}~\bibnamefont{Vogt}},
  \bibinfo{journal}{Phys. Lett.} \textbf{\bibinfo{volume}{B606}},
  \bibinfo{pages}{123} (\bibinfo{year}{2005}), \eprint{hep-ph/0411112}.

\bibitem[{\citenamefont{Vermaseren et~al.}(2005)\citenamefont{Vermaseren, Vogt,
  and Moch}}]{Vermaseren:2005qc}
\bibinfo{author}{\bibfnamefont{J.~A.~M.} \bibnamefont{Vermaseren}},
  \bibinfo{author}{\bibfnamefont{A.}~\bibnamefont{Vogt}}, \bibnamefont{and}
  \bibinfo{author}{\bibfnamefont{S.}~\bibnamefont{Moch}},
  \bibinfo{journal}{Nucl. Phys.} \textbf{\bibinfo{volume}{B724}},
  \bibinfo{pages}{3} (\bibinfo{year}{2005}), \eprint{hep-ph/0504242}.

\bibitem[{\citenamefont{Vogt et~al.}(2006)\citenamefont{Vogt, Moch, and
  Vermaseren}}]{Vogt:2006bt}
\bibinfo{author}{\bibfnamefont{A.}~\bibnamefont{Vogt}},
  \bibinfo{author}{\bibfnamefont{S.}~\bibnamefont{Moch}}, \bibnamefont{and}
  \bibinfo{author}{\bibfnamefont{J.}~\bibnamefont{Vermaseren}},
  \bibinfo{journal}{Nucl. Phys. Proc. Suppl.} \textbf{\bibinfo{volume}{160}},
  \bibinfo{pages}{44} (\bibinfo{year}{2006}), \bibinfo{note}{[,44(2006)]},
  \eprint{hep-ph/0608307}.

\bibitem[{\citenamefont{Moch et~al.}(2008)\citenamefont{Moch, Rogal, and
  Vogt}}]{Moch:2007rq}
\bibinfo{author}{\bibfnamefont{S.}~\bibnamefont{Moch}},
  \bibinfo{author}{\bibfnamefont{M.}~\bibnamefont{Rogal}}, \bibnamefont{and}
  \bibinfo{author}{\bibfnamefont{A.}~\bibnamefont{Vogt}},
  \bibinfo{journal}{Nucl. Phys.} \textbf{\bibinfo{volume}{B790}},
  \bibinfo{pages}{317} (\bibinfo{year}{2008}), \eprint{0708.3731}.

\bibitem[{\citenamefont{Davies et~al.}(2016)\citenamefont{Davies, Vogt, Moch,
  and Vermaseren}}]{Davies:2016ruz}
\bibinfo{author}{\bibfnamefont{J.}~\bibnamefont{Davies}},
  \bibinfo{author}{\bibfnamefont{A.}~\bibnamefont{Vogt}},
  \bibinfo{author}{\bibfnamefont{S.}~\bibnamefont{Moch}}, \bibnamefont{and}
  \bibinfo{author}{\bibfnamefont{J.~A.~M.} \bibnamefont{Vermaseren}},
  \bibinfo{journal}{PoS} \textbf{\bibinfo{volume}{DIS2016}},
  \bibinfo{pages}{059} (\bibinfo{year}{2016}), \eprint{1606.08907}.

\bibitem[{\citenamefont{Salam and Rojo}(2009)}]{Salam:2008qg}
\bibinfo{author}{\bibfnamefont{G.~P.} \bibnamefont{Salam}} \bibnamefont{and}
  \bibinfo{author}{\bibfnamefont{J.}~\bibnamefont{Rojo}},
  \bibinfo{journal}{Comput. Phys. Commun.} \textbf{\bibinfo{volume}{180}},
  \bibinfo{pages}{120} (\bibinfo{year}{2009}), \eprint{0804.3755}.

\bibitem[{\citenamefont{Baglio et~al.}(2011)}]{Arnold:2011wj}
\bibinfo{author}{\bibfnamefont{J.}~\bibnamefont{Baglio}} \bibnamefont{et~al.}
  (\bibinfo{year}{2011}), \eprint{1107.4038}.

\bibitem[{\citenamefont{Butterworth et~al.}(2016)}]{Butterworth:2015oua}
\bibinfo{author}{\bibfnamefont{J.}~\bibnamefont{Butterworth}}
  \bibnamefont{et~al.}, \bibinfo{journal}{J. Phys.}
  \textbf{\bibinfo{volume}{G43}}, \bibinfo{pages}{023001}
  (\bibinfo{year}{2016}), \eprint{1510.03865}.

\bibitem[{\citenamefont{Tanabashi et~al.}(2018)}]{Tanabashi:2018oca}
\bibinfo{author}{\bibfnamefont{M.}~\bibnamefont{Tanabashi}}
  \bibnamefont{et~al.} (\bibinfo{collaboration}{Particle Data Group}),
  \bibinfo{journal}{Phys. Rev.} \textbf{\bibinfo{volume}{D98}},
  \bibinfo{pages}{030001} (\bibinfo{year}{2018}).

\bibitem[{\citenamefont{Alwall et~al.}(2014)\citenamefont{Alwall, Frederix,
  Frixione, Hirschi, Maltoni, Mattelaer, Shao, Stelzer, Torrielli, and
  Zaro}}]{Alwall:2014hca}
\bibinfo{author}{\bibfnamefont{J.}~\bibnamefont{Alwall}},
  \bibinfo{author}{\bibfnamefont{R.}~\bibnamefont{Frederix}},
  \bibinfo{author}{\bibfnamefont{S.}~\bibnamefont{Frixione}},
  \bibinfo{author}{\bibfnamefont{V.}~\bibnamefont{Hirschi}},
  \bibinfo{author}{\bibfnamefont{F.}~\bibnamefont{Maltoni}},
  \bibinfo{author}{\bibfnamefont{O.}~\bibnamefont{Mattelaer}},
  \bibinfo{author}{\bibfnamefont{H.~S.} \bibnamefont{Shao}},
  \bibinfo{author}{\bibfnamefont{T.}~\bibnamefont{Stelzer}},
  \bibinfo{author}{\bibfnamefont{P.}~\bibnamefont{Torrielli}},
  \bibnamefont{and} \bibinfo{author}{\bibfnamefont{M.}~\bibnamefont{Zaro}},
  \bibinfo{journal}{JHEP} \textbf{\bibinfo{volume}{07}}, \bibinfo{pages}{079}
  (\bibinfo{year}{2014}), \eprint{1405.0301}.

\bibitem[{\citenamefont{Actis et~al.}(2017)\citenamefont{Actis, Denner, Hofer,
  Lang, Scharf, and Uccirati}}]{Actis:2016mpe}
\bibinfo{author}{\bibfnamefont{S.}~\bibnamefont{Actis}},
  \bibinfo{author}{\bibfnamefont{A.}~\bibnamefont{Denner}},
  \bibinfo{author}{\bibfnamefont{L.}~\bibnamefont{Hofer}},
  \bibinfo{author}{\bibfnamefont{J.-N.} \bibnamefont{Lang}},
  \bibinfo{author}{\bibfnamefont{A.}~\bibnamefont{Scharf}}, \bibnamefont{and}
  \bibinfo{author}{\bibfnamefont{S.}~\bibnamefont{Uccirati}},
  \bibinfo{journal}{Comput. Phys. Commun.} \textbf{\bibinfo{volume}{214}},
  \bibinfo{pages}{140} (\bibinfo{year}{2017}), \eprint{1605.01090}.

\bibitem[{\citenamefont{Denner et~al.}(2017)\citenamefont{Denner, Dittmaier,
  and Hofer}}]{Denner:2016kdg}
\bibinfo{author}{\bibfnamefont{A.}~\bibnamefont{Denner}},
  \bibinfo{author}{\bibfnamefont{S.}~\bibnamefont{Dittmaier}},
  \bibnamefont{and} \bibinfo{author}{\bibfnamefont{L.}~\bibnamefont{Hofer}},
  \bibinfo{journal}{Comput. Phys. Commun.} \textbf{\bibinfo{volume}{212}},
  \bibinfo{pages}{220} (\bibinfo{year}{2017}), \eprint{1604.06792}.

\bibitem[{\citenamefont{Feger and Pellen}(2015)}]{MoCaNLO}
\bibinfo{author}{\bibfnamefont{R.}~\bibnamefont{Feger}} \bibnamefont{and}
  \bibinfo{author}{\bibfnamefont{M.}~\bibnamefont{Pellen}},
  \emph{\bibinfo{title}{{MoCaNLO: a generic Monte Carlo event generator for NLO
  calculations of hadron-collider processes.}}} (\bibinfo{year}{2015}).

\bibitem[{\citenamefont{Andersen et~al.}(2018)}]{Bendavid:2018nar}
\bibinfo{author}{\bibfnamefont{J.~R.} \bibnamefont{Andersen}}
  \bibnamefont{et~al.}, in \emph{\bibinfo{booktitle}{{10th Les Houches Workshop
  on Physics at TeV Colliders (PhysTeV 2017) Les Houches, France, June 5-23,
  2017}}} (\bibinfo{year}{2018}), \eprint{1803.07977},
  \urlprefix\url{http://lss.fnal.gov/archive/2018/conf/fermilab-conf-18-122-cd-t.pdf}.

\bibitem[{\citenamefont{Denner et~al.}(2015)\citenamefont{Denner, Dittmaier,
  Kallweit, and Mück}}]{Denner:2014cla}
\bibinfo{author}{\bibfnamefont{A.}~\bibnamefont{Denner}},
  \bibinfo{author}{\bibfnamefont{S.}~\bibnamefont{Dittmaier}},
  \bibinfo{author}{\bibfnamefont{S.}~\bibnamefont{Kallweit}}, \bibnamefont{and}
  \bibinfo{author}{\bibfnamefont{A.}~\bibnamefont{Mück}},
  \bibinfo{journal}{Comput. Phys. Commun.} \textbf{\bibinfo{volume}{195}},
  \bibinfo{pages}{161} (\bibinfo{year}{2015}), \eprint{1412.5390}.

\bibitem[{\citenamefont{Biedermann
  et~al.}(2017{\natexlab{a}})\citenamefont{Biedermann, Denner, and
  Pellen}}]{Biedermann:2016yds}
\bibinfo{author}{\bibfnamefont{B.}~\bibnamefont{Biedermann}},
  \bibinfo{author}{\bibfnamefont{A.}~\bibnamefont{Denner}}, \bibnamefont{and}
  \bibinfo{author}{\bibfnamefont{M.}~\bibnamefont{Pellen}},
  \bibinfo{journal}{Phys. Rev. Lett.} \textbf{\bibinfo{volume}{118}},
  \bibinfo{pages}{261801} (\bibinfo{year}{2017}{\natexlab{a}}),
  \eprint{1611.02951}.

\bibitem[{\citenamefont{Biedermann
  et~al.}(2017{\natexlab{b}})\citenamefont{Biedermann, Denner, and
  Pellen}}]{Biedermann:2017bss}
\bibinfo{author}{\bibfnamefont{B.}~\bibnamefont{Biedermann}},
  \bibinfo{author}{\bibfnamefont{A.}~\bibnamefont{Denner}}, \bibnamefont{and}
  \bibinfo{author}{\bibfnamefont{M.}~\bibnamefont{Pellen}},
  \bibinfo{journal}{JHEP} \textbf{\bibinfo{volume}{10}}, \bibinfo{pages}{124}
  (\bibinfo{year}{2017}{\natexlab{b}}), \eprint{1708.00268}.

\bibitem[{\citenamefont{Andersen et~al.}(2008)\citenamefont{Andersen, Binoth,
  Heinrich, and Smillie}}]{Andersen:2007mp}
\bibinfo{author}{\bibfnamefont{J.~R.} \bibnamefont{Andersen}},
  \bibinfo{author}{\bibfnamefont{T.}~\bibnamefont{Binoth}},
  \bibinfo{author}{\bibfnamefont{G.}~\bibnamefont{Heinrich}}, \bibnamefont{and}
  \bibinfo{author}{\bibfnamefont{J.~M.} \bibnamefont{Smillie}},
  \bibinfo{journal}{JHEP} \textbf{\bibinfo{volume}{02}}, \bibinfo{pages}{057}
  (\bibinfo{year}{2008}), \eprint{0709.3513}.

\bibitem[{\citenamefont{Vogt et~al.}(2018)\citenamefont{Vogt, Herzog, Moch,
  Ruijl, Ueda, and Vermaseren}}]{Vogt:2018miu}
\bibinfo{author}{\bibfnamefont{A.}~\bibnamefont{Vogt}},
  \bibinfo{author}{\bibfnamefont{F.}~\bibnamefont{Herzog}},
  \bibinfo{author}{\bibfnamefont{S.}~\bibnamefont{Moch}},
  \bibinfo{author}{\bibfnamefont{B.}~\bibnamefont{Ruijl}},
  \bibinfo{author}{\bibfnamefont{T.}~\bibnamefont{Ueda}}, \bibnamefont{and}
  \bibinfo{author}{\bibfnamefont{J.~A.~M.} \bibnamefont{Vermaseren}},
  \bibinfo{journal}{PoS} \textbf{\bibinfo{volume}{LL2018}},
  \bibinfo{pages}{050} (\bibinfo{year}{2018}), \eprint{1808.08981}.

\bibitem[{\citenamefont{Kaplan and Georgi}(1984)}]{Kaplan:1983fs}
\bibinfo{author}{\bibfnamefont{D.~B.} \bibnamefont{Kaplan}} \bibnamefont{and}
  \bibinfo{author}{\bibfnamefont{H.}~\bibnamefont{Georgi}},
  \bibinfo{journal}{Phys. Lett.} \textbf{\bibinfo{volume}{136B}},
  \bibinfo{pages}{183} (\bibinfo{year}{1984}).

\bibitem[{\citenamefont{Contino et~al.}(2010)\citenamefont{Contino, Grojean,
  Moretti, Piccinini, and Rattazzi}}]{Contino:2010mh}
\bibinfo{author}{\bibfnamefont{R.}~\bibnamefont{Contino}},
  \bibinfo{author}{\bibfnamefont{C.}~\bibnamefont{Grojean}},
  \bibinfo{author}{\bibfnamefont{M.}~\bibnamefont{Moretti}},
  \bibinfo{author}{\bibfnamefont{F.}~\bibnamefont{Piccinini}},
  \bibnamefont{and} \bibinfo{author}{\bibfnamefont{R.}~\bibnamefont{Rattazzi}},
  \bibinfo{journal}{JHEP} \textbf{\bibinfo{volume}{05}}, \bibinfo{pages}{089}
  (\bibinfo{year}{2010}), \eprint{1002.1011}.

\bibitem[{\citenamefont{Agashe et~al.}(2005)\citenamefont{Agashe, Contino, and
  Pomarol}}]{Agashe:2004rs}
\bibinfo{author}{\bibfnamefont{K.}~\bibnamefont{Agashe}},
  \bibinfo{author}{\bibfnamefont{R.}~\bibnamefont{Contino}}, \bibnamefont{and}
  \bibinfo{author}{\bibfnamefont{A.}~\bibnamefont{Pomarol}},
  \bibinfo{journal}{Nucl. Phys.} \textbf{\bibinfo{volume}{B719}},
  \bibinfo{pages}{165} (\bibinfo{year}{2005}), \eprint{hep-ph/0412089}.

\bibitem[{\citenamefont{Contino et~al.}(2007)\citenamefont{Contino, Da~Rold,
  and Pomarol}}]{Contino:2006qr}
\bibinfo{author}{\bibfnamefont{R.}~\bibnamefont{Contino}},
  \bibinfo{author}{\bibfnamefont{L.}~\bibnamefont{Da~Rold}}, \bibnamefont{and}
  \bibinfo{author}{\bibfnamefont{A.}~\bibnamefont{Pomarol}},
  \bibinfo{journal}{Phys. Rev.} \textbf{\bibinfo{volume}{D75}},
  \bibinfo{pages}{055014} (\bibinfo{year}{2007}), \eprint{hep-ph/0612048}.

\bibitem[{\citenamefont{Goldberger et~al.}(2008)\citenamefont{Goldberger,
  Grinstein, and Skiba}}]{Goldberger:2008zz}
\bibinfo{author}{\bibfnamefont{W.~D.} \bibnamefont{Goldberger}},
  \bibinfo{author}{\bibfnamefont{B.}~\bibnamefont{Grinstein}},
  \bibnamefont{and} \bibinfo{author}{\bibfnamefont{W.}~\bibnamefont{Skiba}},
  \bibinfo{journal}{Phys. Rev. Lett.} \textbf{\bibinfo{volume}{100}},
  \bibinfo{pages}{111802} (\bibinfo{year}{2008}), \eprint{0708.1463}.

\bibitem[{\citenamefont{\texttt{proVBFH-incl} v2.0.0}()}]{proVBFH}
\bibinfo{author}{\bibnamefont{\texttt{proVBFH-incl} v2.0.0}},
  \bibinfo{note}{{\url{http://provbfh.hepforge.org/}}}.

\end{thebibliography}

\end{document}